\documentclass[11pt,a4paper,reqno]{amsart}
\usepackage[margin=1in]{geometry}
\usepackage[T1]{fontenc}
\usepackage[utf8]{inputenc}
\usepackage[british]{babel}
\usepackage{amsmath,amssymb,amsthm}
\usepackage{bm}
\usepackage{xcolor}
\usepackage{adjustbox}
\usepackage{tikz-cd}
\usepackage[hidelinks]{hyperref}
\usepackage[capitalise,noabbrev]{cleveref}

\newtheorem{theorem}{Theorem}[section]
\newtheorem{lemma}[theorem]{Lemma}
\newtheorem{proposition}[theorem]{Proposition}
\newtheorem{corollary}[theorem]{Corollary}
\theoremstyle{definition}
\newtheorem{definition}[theorem]{Definition}
\newtheorem{example}[theorem]{Example}
\theoremstyle{remark}
\newtheorem{remark}[theorem]{Remark}
\newenvironment{nestedproof}{\begin{proof}}{\end{proof}}

\title{k-partite quantum homomorphisms}
\author{Amin Karamlou}
\makeatletter
\let\original@setauthors\@setauthors
\renewcommand{\@setauthors}{%
  \original@setauthors
  \begin{center}
    \footnotesize
    Department of Computer Science, University College London\\
    \texttt{a.karamlou@ucl.ac.uk}
  \end{center}
}
\makeatother

\newcommand{\N}{\mathbb{N}}
\newcommand{\C}{\mathbb{C}}

\newcommand{\PVM}{\operatorname{PVM}}

\newcommand{\supp}{\operatorname{supp}}
\newcommand{\ket}[1]{|#1\rangle}
\newcommand{\bra}[1]{\langle#1|}
\newcommand{\HH}{\mathbf{H}}

\newcommand{\Kl}{\mathrm{Kl}}
\newcommand{\id}{\mathrm{id}}

\DeclareMathOperator{\ar}{ar}

\newcommand{\Proj}{\operatorname{Proj}}

\newcommand{\cat}[1]{\mathbf{#1}}
\newcommand{\As}{\mathcal{A}}
\newcommand{\Bs}{\mathcal{B}}
\newcommand{\Cs}{\mathcal{C}}

\newcommand{\Rsig}{\cat{R(\sigma)}}
\newcommand{\ksig}{\cat{R^k(\sigma)}}
\newcommand{\cfk}{\langle k \rangle}
\newcommand{\Syst}{M^t_k}
\newcommand{\Sysco}{{M}^{co}_k}

\newcommand{\catSet}{\cat{Set}}

\begin{document}

\begin{abstract}
$k$-partite XOR games and linear BCS games are two of the earliest studied classes of non-local games. Both arise from systems of linear equations over $\mathbb{Z}_2$, thus providing distinct quantum relaxations of the classical constraint satisfaction problem, $\mathsf{CSP(\mathcal{XOR})}$.

Linear BCS games have been generalised to synchronous homomorphism games on arbitrary relational structures, which provide quantum relaxations for every classical CSP. Moreover, the resulting notion of quantum homomorphism admits compositional semantics via the quantum monad. Studying this monad has significantly advanced our understanding of (synchronous) quantum advantage in CSPs. No analogous generalisation of $k$-partite XOR games has previously been developed.

We introduce $k$-homomorphism games, extending $k$-XOR games from linear systems to arbitrary relational structures. This yields a new $k$-partite model for studying quantum CSPs and quantum homomorphisms, which we characterise using a $k$-partite variant of the quantum monad. We then study the occurrence of $k$-partite quantum advantage for CSPs, and compare this with the occurrence of synchronous quantum advantage. More specifically, we show that:

\begin{itemize}
    \item There exist CSP instances which have quantum advantage in the $k$-partite model but not in the synchronous model, and vice-versa.
    \item There exist instances of NP-complete CSPs which admit $k$-partite quantum advantage.
    \item No bounded-width CSP template admits $k$-partite quantum advantage. 
\end{itemize}
\end{abstract}

\maketitle
\section{Introduction}

\subsection{Games, constraints, and homomorphisms}

A striking feature of quantum mechanics is that parties based in spatially separated laboratories can, with the aid of shared entanglement, produce correlations that cannot be explained by any local hidden-variable theory. This phenomenon, known as (quantum) non-locality, was first highlighted by Bell's theorem \cite{bell1964einstein} and has since become one of the defining distinctions between classical and quantum information processing. While these correlations may initially appear paradoxical, they have been confirmed by increasingly sophisticated Bell experiments \cite{Freedman1972experimental,Aspect1982experimental}, culminating in ``loophole-free'' tests \cite{Hensen2015experiment} that firmly establish non-locality as an experimentally verified feature of nature.

Inspired by the importance of interactive proof systems in theoretical computer science, Cleve, Høyer, Toner and Watrous \cite{cleve2004consequences} introduced the framework of non-local games as a setting for studying non-locality. These games are played between a verifier (or referee) and $k$ players via the following protocol:

\begin{enumerate}
    \item The verifier sends a question to each player.
    \item Each player responds with an answer.
    \item The verifier evaluates the questions and answers against a pre-defined predicate to determine if the players win or lose. 
\end{enumerate}

Players may agree on a strategy for the game beforehand but once the game begins they are not allowed to communicate. Classical strategies allow the players to correlate their responses using shared randomness established before the game, whereas quantum strategies additionally permit the players to share a quantum state. The existence of games in which quantum strategies outperform every classical strategy demonstrates the computational power of non-locality.

Many non-local games arise from instances of constraint satisfaction problems (CSPs). A CSP instance asks whether a finite collection of constraints admits a satisfying assignment. Every such instance can be reformulated as the question of whether a homomorphism exists between relational structures:

\begin{description}
    \item[$\mathsf{CSP}$]: Given two relational structures $\As, \Bs$ decide if $\As \rightarrow \Bs$.
\end{description}

Also of importance is the so-called fixed-template variant of the above problem.

\begin{description}
    \item[$\mathsf{CSP(\Bs)}$]: Given a relational structure $\As$ decide if $\As \rightarrow \Bs$.
\end{description}

Formulating constraint satisfaction as a homomorphism problem provides a powerful framework for describing a wide range of decision problems, and connects their study to universal algebra \cite{bulatov2005classifying}, finite model theory \cite{kolaitis2008logical}, and category theory \cite{hadek2025categorical}. This viewpoint has proved extremely useful, most notably leading to two independent positive resolutions of the Feder-Vardi dichotomy conjecture \cite{feder1998conjecture} by Bulatov \cite{bulatov2017dichotomy} and Zhuk \cite{zhuk2020dichotomy}.

% CSPs offer a powerful framework A remarkable range of important decision problems, such as graph colouring, linear-equation solving, and variations of satisfiability solving can be phrased as constraint satisfaction problems (CSPs).

Two of the earliest studied classes of non-local games are $k$-partite XOR games \cite{cleve2004consequences, bene_watts_2021identifiying} and linear binary constraint system (BCS) games \cite{cleve2014characterization}. Both originate from systems of linear equations over $\mathbb{Z}_2$, and therefore provide alternative quantum relaxations of the same classical decision problem $\mathsf{CSP(\mathcal{XOR})}$. Despite this common origin, the two constructions are fundamentally different. The $k$-XOR game is a $k$-partite non-local game in which the variables must be partitioned among the players, with exactly one variable corresponding to each player being involved in each equation. A Linear BCS game on the other hand, is a two-player non-local game in which each player is given an entire clause as a question. 

% Linear BCS games are synchronous, meaning that Alice and Bob have the same question and answer set, and must provide the same answer when receiving the same question in order to win. $k$-XOR games are not (in general) synchronous. 

This distinction is reflected in the quantum phenomena that each construction captures. We say that a non-local game exhibits pseudotelepathy if it admits a perfect quantum strategy but no perfect classical strategy. The Greenberger--Horne--Zeilinger (GHZ) game \cite{greenberger1989going} is a well-known pseudotelepathic $3$-XOR game, but a theorem of Arkhipov \cite{arkhipov2012extending} shows that the corresponding linear system has no perfect quantum strategy in the BCS formulation. Conversely, the Mermin--Peres magic square game \cite{mermin1993hidden, peres1990incompatible} is a pseudotelepathic linear BCS game, yet we will show in this paper that its system of linear equations does not admit a perfect quantum strategy when formulated as a $k$-XOR game. A further curious difference between the frameworks is that for several classes of $k$-XOR games, such as symmetric $k$-XOR games, and $3$-XOR games there are efficient algorithms which determine if a perfect quantum strategy exists \cite{watts20233xor, benewatts2019algorithms}. The corresponding question for linear BCS games is known to be undecidable \cite{slofstra2019set, slofstra2019tsirelson}. 

In \cite{abramsky2017quantum}, linear BCS games are generalised to synchronous\footnote{A synchronous non-local game is a two-player game where both players have the same input and output set, and where they must respond with the same output when given the same input in order to win. Strictly speaking, the formulation of the homomorphism game given in \cite{abramsky2017quantum} is not synchronous. Their game is however equivalent to the synchronous version we consider here \cite{kim2018synchronous, karamlou2025composing}.} homomorphism games on relational structures:

\begin{definition}
Consider relational structures $\As, \Bs$. The synchronous $(\mathcal{A}, \mathcal{B})$-homomorphism game denoted by $G^s(\As, \Bs)$ is played as  follows:
\begin{enumerate}
    \item Verifier sends player 1 (Alice) and player 2 (Bob) tuples $\vec{x}_a, \vec{x}_b \in A^n$ respectively. 
    \item Alice and Bob return tuples $\vec{y}_a, \vec{y}_b \in B^n$ respectively.
    \item Alice and Bob win the game if:
\begin{itemize}
        \item $x_{a,i} = x_{b,j} \implies y_{a,i} = {y}_{b, j}$.
	\item $\forall R \in \sigma: \vec{x}_a \in R^\As \implies \vec{y}_a \in R^\Bs$ and $\vec{x}_b \in R^\As \implies \vec{y}_b \in R^\Bs$.
 \end{itemize}
\end{enumerate}
\end{definition}

This game is closely connected to a substantial body of work in quantum information, including the theory of synchronous games and their associated *-algebras (see the recent survey \cite{harris2025games}), state-independent contextuality \cite{abramsky2017quantum}, quantum graph theory \cite{mancinska2016quantum, atserias2019quantum, mancinska2018oddities, mancinska2020quantum}, quantum groups \cite{lupini2020nonlocal}, and quantum logic \cite{dawar2026complexity}. Of relevance to the present work is the observation that perfect (finite-dimensional) quantum strategies in this game admit a categorical characterisation as Kleisli morphisms of a (graded) quantum monad $Q_d$:

\begin{theorem}[\cite{abramsky2017quantum}]
\label{theorem:synch-quantum-monad} $\As \rightarrow Q_d \Bs$ if and only if  $G^s(\As, \Bs)$ has a perfect finite-dimensional quantum strategy.
\end{theorem}

% This theorem allows us to formulate synchronous quantum relaxations of the decision problems $\mathsf{CSP}$ and $\mathsf{CSP(\Bs)}$.

% \begin{description}
%     \item[$\mathsf{S-QuCSP}$]: Given two relational structures $\As, \Bs$ decide if $\As \rightarrow Q_{\mathbf{H}} \Bs$.
% \end{description}

% \begin{description}
%     \item[$\mathsf{S-QuCSP(\Bs)}$]: Given a relational structures $\As$ decide if $\As \rightarrow Q_{\mathbf{H}} \Bs$.
% \end{description}

In what follows we will refer to maps $\As \rightarrow Q_{d} \Bs$ as synchronous quantum homomorphisms. This concept opens up the possibility of studying quantum advantage for CSPs:

\begin{definition}
\label{def:synch-advantage}

A CSP instance $(\As, \Bs)$ admits (pseudotelepathic) synchronous quantum advantage if $\As \not \rightarrow \Bs$ and $\As \rightarrow Q_d \Bs$. We say that $\mathsf{CSP({\mathcal{B})}}$ admits quantum advantage if there exists some $\As$ for which $(\As, \Bs)$ admits quantum advantage. 
\end{definition}
 
In \cite{ciardo2024quantum} the quantum monad is combined with
techniques from universal algebra to obtain conditions under
which quantum advantage occurs. A CSP template
$\Bs$ has \emph{bounded width} if there is a fixed level of
local consistency \cite{feder1998conjecture} sufficient to decide $\mathsf{CSP(\Bs)}$.
Informally, local consistency algorithms reason about partial
assignments to small sets of variables and their compatibility,
rather than searching directly for a global satisfying assignment.
One of the main results of \cite{ciardo2024quantum} establishes
the following connection between bounded width and quantum
advantage.

\begin{theorem}[\cite{ciardo2024quantum}]
\label{thm:ciardo-width}
If $\mathsf{CSP(\Bs)}$ admits synchronous quantum advantage it has unbounded width. 
\end{theorem}

\subsection{Contributions}

The purpose of this paper is to generalise $k$-XOR games to $k$-homomorphism games in a manner that parallels the developments described above for linear BCS games. Our main contributions are as follows:

\begin{enumerate}
    \item In Section \ref{sec:games} we introduce $k$-partite homomorphism games, a generalisation of $k$-partite XOR games to arbitrary relational structures. 
    
    % We show that this class of games is complete, in the sense that every $k$-partite non-local game is equivalent to a $k$-homomorphism game. 

    \item In Section \ref{sec:classical} we explore perfect classical strategies, and show that these exist precisely when classical homomorphisms exist. Composition of strategies in the $k$-partite model turns out to be more nuanced than the synchronous model. We identify a category of $k$-partite relational structures $\ksig$, and show that this is the correct setting for composing classical strategies.
    
    % There is a forgetful functor $U:\ksig \rightarrow \Rsig$ which admits a right adjoint. This right adjoint provides a cofree construction on relational structures, which allows us to identify homomorphisms of relational structures within the more structured category of k-partite structures.

    \item In Section \ref{sec:quantum} we explore quantum strategies. We introduce a $k$-partite (graded) quantum monad on $\ksig$, whose Kleisli morphisms characterise perfect quantum strategies for the new games. We are able to explain both commuting operator and tensor product strategies using variations of this construction.

    \item In Section \ref{sec:beyond-xor} we define the concept of $k$-partite quantum advantage for CSPs. We show that there exists a CSP template other than $\mathcal{XOR}$, in fact an NP-complete one,  which admit this form of advantage.

    \item In Section \ref{sec:sync-vs-k-partite} we separate synchronous and $k$-partite quantum advantage by showing that there exist CSP instances which admit synchronous advantage but not $k$-partite advantage, and vice-versa.

    \item In Section \ref{sec:bounded-width} we prove a $k$-partite version of theorem \ref{thm:ciardo-width}. Namely, that a CSP template which admits $k$-partite quantum advantage must have unbounded width. 

\end{enumerate}

\section{Background}
\label{sec:background}
\label{app:background}
\subsection{Category theory}

We assume familiarity with the definitions of category, functor, and natural transformation. These can be found in standard introductory texts on category theory \cite{mac1971categories}. 

\begin{definition}
	Given a monoid $(M,\cdot,1)$, an \textbf{${M}$-\emph{graded monad}} $(T_m,\mu_{m,m'},\eta)$ over a category $\cat{C}$ consists of:
    \begin{itemize}
    \item A family of endofunctors $T_m: \cat{C} \rightarrow \cat{C}$ for each $m \in M$;
    \item A family of natural transformations $\mu_{m,m'}: T_m T_{m'} \Rightarrow T_{m\cdot m'}$, for each $m,m' \in M$. These are referred to as the \textbf{multiplication} operations.
    \item A natural transformation $\eta: \id_{\cat{C}} \Rightarrow T_1$. This is referred to as the \textbf{unit} operation.
    \end{itemize}
      These must satisfy the following commutativity conditions:

\medskip
\[
\begin{tikzcd}[column sep=huge, row sep=large]
T_m T_{m'}T_{m''} A
  \arrow[r, "{T_m((\mu_{m',m''})_A)}"]
  \arrow[d, "{(\mu_{m,m'})_{T_{m''}A}}"']
& T_m T_{m'\cdot m''} A
  \arrow[d, "{(\mu_{m,m'\cdot m''})_A}"]
\\
T_{m\cdot m'} T_{m''} A
  \arrow[r, "{(\mu_{m\cdot m',m''})_A}"']
& T_{m\cdot m'\cdot m''} A
\end{tikzcd}
\]
\[
\begin{tikzcd}[column sep=large, row sep=large]
& T_1 T_m A \arrow[dr, "{(\mu_{1,m})_A}"] & \\
T_m A \arrow[ur, "{\eta_{T_m A}}"]
      \arrow[rr, "{\id_{T_m A}}"'] && T_m A
\end{tikzcd}
\qquad
\begin{tikzcd}[column sep=large, row sep=large]
& T_m T_1 A \arrow[dr, "{(\mu_{m,1})_A}"] & \\
T_m A \arrow[ur, "{T_m(\eta_A)}"]
      \arrow[rr, "{\id_{T_m A}}"'] && T_m A
\end{tikzcd}
\]
More generally, a graded monad is a (lax) monoidal functor $(\cat{M}, \otimes, I) \rightarrow ([\cat{C}, \cat{C}], \circ, \id[\cat{C}])$ from a monoidal category to the category of endofunctors over $\cat{C}$\cite{fujii2016towards}.

\end{definition}

\subsection{Model theory}

A finite relational signature $\sigma$ is a set of relation symbols $\sigma = \{R_1, \ldots, R_n\}$ where each symbol $R \in \sigma$ has an associated non-negative integer $ar(R)$ called its arity. A relational structure $\As$ with signature $\sigma$ (also known as a $\sigma$-structure) is a tuple $(A, R_1^\As, \ldots, R_n^\As)$ where $A$ is the underlying set of $\As$ known as its \emph{universe} and for each $R \in \sigma, R^\As \subseteq A^{ar(R)}$ is a relation of arity $ar(R)$. We shall use the calligraphic font to distinguish a relational structure $\As$ from its universe $A$. A relational structure is said to be finite if its universe is finite. (Finite) relational structures model many important data structures in computer science. For instance, if $\sigma$ consists of a single relational symbol of arity $2$ then $\sigma$-structures are precisely directed graphs. By varying the symbols in our signature\footnote{We may also occasionally have to impose additional properties such as symmetry or irreflexivity on our relations to capture data structures such as undirected graphs and acyclic graphs.} we can capture other data structures such as multi-graphs, coloured graphs, strings, finite transition systems, and more. Relational structures thus provide us with a uniform language to simultaneously prove results about all these data structures. 

 Let $\As$ and $\Bs$ be two relational structures over the same signature $\sigma$.
 
 A function $f: A \rightarrow B$ is said to \emph{preserve} a relation $R \in \sigma$ with arity $n$ if

$$\forall \vec{a} = (a_1, \ldots, a_n) \in A^n: \vec{a} \in R^\As \implies \vec{f}(\vec{a}) = (f(a_1), \ldots, f(a_n)) \in R^\Bs,$$

where we have used the vector notation $\vec{a}$ to denote tuples. $f$ is said to be a homomorphism if it preserves every $R \in \sigma$. The category $\Rsig$ has $\sigma$-structures as objects and $\sigma$-structure homomorphisms as morphisms.

\subsection{Quantum information}

We associate to any isolated physical system a Hilbert space $\mathbf{H}$ known as its \emph{state space}. A \emph{pure quantum state} is a vector of unit norm in $\mathbf{H}$. The intuition is that each such vector represents a possible state of the physical system.

We recall that a bounded linear operator $N$ on $\pmb{H}$ is said to be:

\begin{itemize}
    \item \emph{Positive semi-definite} if $\forall u \in H, \langle N u, u\rangle\geq {0}$. In this case, we write $N \geq \pmb{0}$. Likewise, the term \emph{positive definite} is used if we replace the $\geq$ condition with $>$.
    \item Hermitian if $N = N^\dagger$.
    \item An (orthogonal) projector if $N = N^2 = N^\dagger$.
\end{itemize}

We write $\Proj(\mathbf{H})$ for the set of projectors on $\mathbf{H}$. We refer to two projectors $M, N$ as being orthogonal to each other if $MN = \pmb{0}$. We write $M \leq N$ iff $N - M$ is positive semi-definite. $M$ and $N$ \emph{commute} iff $M N - N M = \pmb{0}$. The product of two projectors is a projector iff they commute. For any finite family of projectors $\{P_i\}_i$ it holds that $\sum_i P_i \leq I_{\pmb{H}}$ iff $P_i P_j = \pmb{0}$ whenever $i \neq j$. 

To observe a property of a physical system we must perform a measurement on it. In this paper we will restrict attention to \emph{projection valued measurements} (PVM). A PVM is a function  $P: O \rightarrow \Proj(\mathbf{H})$ which satisfies $\sum_{o \in O} P(o) = I_\mathbf{H}$. Here, the elements of $O$ represent the possible outcomes of a measurement. We write $\PVM(\mathbf{H}, O)$ for the set of finitely-supported PVMs in $\mathbf{H}$ with outcomes in the set $O$.

The outcome of a measurement in quantum theory is probabilistic. If a PVM $P: O \rightarrow \mathcal{B}^+(\mathbf{H})$ is performed on a pure state $\psi \in \mathbf{H}$ outcome $o \in O$ will be observed with probability ${\psi}^\dagger P(o)\psi$.

A key feature of quantum theory (one which is in fact necessary for quantum advantage \cite{uola2014joint}) is that not all measurements can be performed jointly. Joint measurements are precisely those that can be performed in any order without affecting the outcome probabilities of the individual measurements. We will write $P_1 \odot P_2$ whenever $P_1$ and $P_2$ are jointly measurable. We say that $P_1$ and $P_2$ commute if every projector appearing in $P_1$ commutes with every projector appearing in $P_2$.

\begin{lemma}[\cite{heinosaari2008notes}]
\label{lemma:PVM-commeasurability}
For a set of PVMs $\{P_i\}_i$ the following are equivalent:

\begin{enumerate}
    \item The $P_i$ are pairwise commuting.
    \item There exists a unique joint measurement $P$ which is projective and defined as $P(o_1, \ldots, o_n) = \prod_{i \in [n]} P_i(o_i)$.
\end{enumerate}
\end{lemma}

% Given Hilbert spaces $\mathbf{H}_1,\ldots,\mathbf{H}_k$, their tensor
% product
% \[
%     \mathbf{H}_1\otimes\cdots\otimes\mathbf{H}_k
% \]
% is the state space associated with the composite system. If
% $T_i\in B(\mathbf{H}_i)$, their tensor product acts on elementary
% tensors by
% \[
%     (T_1\otimes\cdots\otimes T_k)
%     (v_1\otimes\cdots\otimes v_k)
%     :=
%     T_1v_1\otimes\cdots\otimes T_kv_k,
% \]
% and extends linearly and continuously to the full tensor-product
% space.

% In particular, an operator $T\in B(\mathbf{H}_i)$ acting locally on
% the $i$th subsystem is represented on the composite system by
% \[
%     I_{\mathbf{H}_1}\otimes\cdots\otimes
%     I_{\mathbf{H}_{i-1}}\otimes T\otimes
%     I_{\mathbf{H}_{i+1}}\otimes\cdots\otimes
%     I_{\mathbf{H}_k}.
% \]
% Operators acting on distinct tensor factors commute.

% We write $B(\mathbf{H})$ for the algebra of bounded linear operators
% $T:\mathbf{H}\to\mathbf{H}$, equipped with operator composition,
% the adjoint operation $T\mapsto T^\dagger$, and the operator norm
% \[
%     \|T\| := \sup_{\|v\|=1}\|Tv\|.
% \]

A \emph{$C^*$-algebra} is a complex Banach algebra $\mathfrak A$
equipped with an involution $a\mapsto a^*$ satisfying
\[
    \|a^*a\|=\|a\|^2
    \qquad
    \text{for every }a\in\mathfrak A.
\]
Here an involution satisfies
\[
    (a^*)^*=a,\qquad
    (ab)^*=b^*a^*,\qquad
    (\lambda a+b)^*=\overline{\lambda}a^*+b^*
\]
for all $a,b\in\mathfrak A$ and $\lambda\in\mathbb C$.
The canonical example is $B(\mathbf H)$, where the involution is
the adjoint $T\mapsto T^\dagger$.

A $C^*$-algebra $\mathfrak A$ is \emph{unital} if it contains an
element $1_{\mathfrak A}$ satisfying
\[
    1_{\mathfrak A}a=a1_{\mathfrak A}=a
    \qquad
    \text{for every }a\in\mathfrak A.
\]
An element $p\in\mathfrak A$ is a \emph{projection} if
\[
    p=p^*=p^2.
\]

Let $\mathfrak A$ be a unital $C^*$-algebra and let $O$ be a set.
A \emph{projective measurement} on $\mathfrak A$ with outcome set
$O$ is a finitely supported map
\[
    P:O\longrightarrow\mathfrak A
\]
such that each $P(o)$ is a projection,
\[
    P(o)P(o')=0
    \qquad\text{whenever }o\neq o',
\]
and
\[
    \sum_{o\in O}P(o)=1_{\mathfrak A}.
\]
We write $\PVM(\mathfrak A,O)$ for the set of all such projective
measurements. 
Two $C^*$-subalgebras $\mathfrak A,\mathfrak B\subseteq\mathfrak C$
of a common $C^*$-algebra $\mathfrak C$ are said to \emph{commute} if
\[
    ab=ba
    \qquad
    \text{for every }a\in\mathfrak A,\ b\in\mathfrak B.
\]
A family $(\mathfrak A_i)_{i\in[k]}$ of $C^*$-subalgebras of
$\mathfrak C$ is \emph{pairwise commuting} if
$\mathfrak A_i$ and $\mathfrak A_j$ commute whenever $i\neq j$.

Let $\mathfrak A$ and $\mathfrak B$ be $C^*$-algebras. Their
\emph{algebraic tensor product} $\mathfrak A\odot\mathfrak B$
consists of finite sums
\[
    \sum_{i=1}^n a_i\otimes b_i,
    \qquad
    a_i\in\mathfrak A,\quad b_i\in\mathfrak B,
\]
with multiplication and involution determined by
\[
    (a\otimes b)(a'\otimes b')
    =aa'\otimes bb',
    \qquad
    (a\otimes b)^*=a^*\otimes b^*,
\]
and extended linearly.

The \emph{minimal tensor product}
$\mathfrak A\otimes_{\min}\mathfrak B$ is the completion of
$\mathfrak A\odot\mathfrak B$ with respect to the minimal
$C^*$-norm. Concretely, if
\[
    \pi:\mathfrak A\to B(\mathbf H),
    \qquad
    \rho:\mathfrak B\to B(\mathbf K)
\]
are faithful $*$-representations, then
\[
    \|x\|_{\min}
    :=
    \|(\pi\otimes\rho)(x)\|_{B(\mathbf H\otimes\mathbf K)},
    \qquad
    x\in\mathfrak A\odot\mathfrak B,
\]
and this norm is independent of the choice of faithful
representations. Thus $\mathfrak A\otimes_{\min}\mathfrak B$ may be
realised as the operator-norm closure of
$(\pi\otimes\rho)(\mathfrak A\odot\mathfrak B)$ in
$B(\mathbf H\otimes\mathbf K)$.

% \subsection{non-local games}

% There are several classes of strategies for generating such correlations. We explain the ones studied here below. 

% \amin{There should probably be an argument added about how POVMs don't buy you anything more than PVMs.}

% A two-player non-local game is said to be \emph{synchronous} if Alice and Bob have the same question set $X$, the same answer set $Y$, and the following equation is satisfied:
% \begin{align*}
% 	V(y, y', x, x) = 0 \quad   \forall x \in X, y \neq y' \in Y
% \end{align*}

% This means that given the same inputs, Alice and Bob must always respond with the same outputs to win the game.

% The main theorem of \cite{abramsky2017quantum} states that there exists a graded quantum monad $Q_{\mathbf{H}}: \Rsig \rightarrow \Rsig$ on the category of relational structures whose Kleisli morphisms characterise perfect quantum strategies in the synchronous homomorphism game. 

% One of our goals is to derive a variant of this result for $k$-partite quantum homomorphisms.

\section{k-player homomorphism game}
\label{sec:games}

Our goal in this section is to generalise $k$-XOR games to games for arbitrary CSP instances. Let us begin by formalising our definition of $k$-partite non-local games.

\begin{definition}
    A $k$-partite non-local game is defined as $G=(X_1, \ldots, X_k, Y_1, \ldots, Y_k, \pi, V)$ where the $X_i$ are finite input sets, the $Y_i$ are finite output sets, $\pi$ is a probability distribution on $\Pi_{i\in [k]} X_i$, and $V: \Pi_{i\in [k]} X_i \times \Pi_{i\in [k]} Y_i \rightarrow \{0, 1\}$ is called the winning predicate. We will assume that $\pi$ has full support. The game proceeds according to the following protocol:
\begin{enumerate}
\item The verifier samples a tuple $\vec{x} = (x_1, \ldots, x_k) \in \Pi_{i\in [k]} X_i$ using $\pi$ and sends $x_i$ to player $i$.
\item Without communicating, each player responds with $y_i \in Y_i$.
\item The players win the game if $V(\vec{x}, \vec{y}) = 1$.
\end{enumerate}
\end{definition}
We focus on cases where the game consists of only one round of the above protocol. The goal of the players is to maximise their winning probability. To achieve this goal they are allowed to agree on a fixed strategy $S$ for the game beforehand. A correlation\footnote{Also known as a behaviour or empirical model.} $p(\vec{y} | \vec{x})$ represents the joint conditional probability of players responding with $\vec{y}$ on inputs $\vec{x}$. We are concerned only with perfect strategies, where the game is won with probability 1. In this case we must have $V(\vec{x},\vec{y}) = 0 \implies p(\vec{y} | \vec{x}) = 0$ and we may assume that $\pi$ always has full support. 

We can now recall the definition of $k$-XOR games from \cite{bene_watts_2021identifiying}.

\begin{definition}

Let $Q$ be a finite set of questions together with a $k$-partition $
Q = Q_1 \sqcup \cdots \sqcup Q_k.$ A $k$-partite clause is a pair $(q,s)$ where
$q \in Q_1 \times \cdots \times Q_k$ and $s\in\{-1,1\}$.

Given a finite set of $k$-partite clauses \(M\), the associated \(k\)-XOR game is played as follows.

\begin{enumerate}
    \item The verifier selects a clause \(c_i=(q_i,s_i)\) from \(M\).
    \item For each \(\alpha\in[k]\), the verifier sends the question
    $q_i^{(\alpha)}$ to the \(\alpha\)-th player.
    \item Each player returns an output
    $a_\alpha \in \{-1,1\}.$
    \item The players win if $\prod_{\alpha=1}^{k} a_\alpha = s_i.$
\end{enumerate}
\end{definition}

Many well-studied games such as CHSH \cite{clauser1969proposed} and GHZ \cite{greenberger1989going} are examples of $k$-XOR games.

% \begin{example}
% \label{ex:GHZ}
% Consider the system of equations
% \[
% \begin{aligned}
% &c_1 : a_0 \oplus b_0 \oplus c_0 = 0, \quad &c_2: a_0 \oplus b_1 \oplus c_1 = 1,\\
% &c_3 : a_1 \oplus b_0 \oplus c_1 = 1, \quad &c_4 : a_1 \oplus b_1 \oplus c_0 = 1.
% \end{aligned}
% \]

% The corresponding $3$-XOR game is precisely the GHZ game.
% \end{example}

A brief remark about the $k$-partiteness condition is in order. Due to this condition there are instances of $\mathsf{CSP}(\mathcal{XOR}_k)$ which cannot be naturally encoded as a $k$-XOR game for the same value of $k$. We provide a simple example.

\begin{example}
Consider the instance of $\mathsf{CSP}(\mathcal{XOR}_2)$ given by the equations
\[
\begin{aligned}
c_1: a \oplus b &= 0, \quad c_2: b \oplus c = 0, \quad c_3: a \oplus c = 1.
\end{aligned}
\]
Suppose we wish to encode this instance as a $2$-XOR game. We must find a partition of the question set $Q = \{a, b, c\}$ into $Q_1 \sqcup Q_2$ with respect to which every clause is $2$-partite. $c_1$ forces us to map $a$ and $b$ into different partitions. Assume without loss of generality that we pick $a \mapsto Q_1$ and $b \mapsto Q_2$. Then $c_2$ forces $c \mapsto Q_1$. Now the $2$-partition does not respect $c_3$ since both $a$ and $c$ are mapped to $Q_1$.
\end{example}

The trouble with the above example is that the clauses force the variables into three different partitions. Thus, in order to encode it we would need to have a third player. This observation guides our generalisation which we now describe.

% In general, any instance of CSP$(\mathrm{k-XOR}$) can be encoded by a $k$-XOR game for some value of $k'$. 

We write $G(\mathcal{\As})$ to denote the Gaifman graph of the structure $\mathcal{A}$. This is the graph with universe $A$ where an edge $a\sim a'$ exists whenever $a \neq a'$ occur together in some tuple of some relation of $\mathcal{A}$.

\begin{definition}
\label{def:game}
Fix a $k \in \mathbb{N}$. Let $\As$ and $\Bs$ be finite relational structures, and let $f: G(\As) \rightarrow K_k$ be a $k$-colouring of $G(\As)$. The $k$-partite homomorphism game $\mathbb{G}_f(\As, \Bs)$ is played as follows:

\begin{enumerate}
\item The verifier selects a relation $R \in \sigma$ and a tuple $\vec{a} \in R^\mathcal{A}$.
\item The verifier sends the question
$a_i$ to player $f(a_i)$;
\item Each player returns an output $b_i \in B$;
\item The players win if $\vec{b} \in R^\Bs$;
\end{enumerate}
\end{definition}

This game is a generalisation of the $k$-XOR game. More specifically, let $\sigma_k:=\{R_0,R_1\}$ be the signature consisting of two
$k$-ary relation symbols. We define the $\sigma_k$-structure
$	\mathcal{XOR}_k$ to have universe $\{0,1\}$ and relations
\[
R_\beta^{	\mathcal{XOR}_k}
:=
\left\{
(b_1,\ldots,b_k)\in\{0,1\}^k
:
b_1\oplus\cdots\oplus b_k=\beta
\right\},
\qquad \beta\in\{0,1\}.
\]

Now let $M$ be a set of $k$-partite clauses.
Identifying $(-1)^\beta\in\{-1,1\}$ with $\beta\in\{0,1\}$, define
the $\sigma_k$-structure $\mathcal A_M$ to have universe $Q$ and
relations
\[
R_\beta^{\mathcal A_M}
:=
\left\{
(q_1,\ldots,q_k)\in Q_1\times\cdots\times Q_k
:
\bigl((q_1,\ldots,q_k),(-1)^\beta\bigr)\in M
\right\}.
\]
The given $k$-partition of $Q$ induces a colouring $p$ given by
$p(q)=i \Longleftrightarrow q\in Q_i.$
This colouring is proper, since each tuple of $\mathcal A_M$
contains exactly one element from each $Q_i$.

\begin{proposition}
    
\label{prop:kxor-khom}
Up to the identification $\{-1,1\}\cong\{0,1\}$ given by
$(-1)^b\leftrightarrow b$, the $k$-XOR game associated with $M$ is
exactly the $k$-homomorphism game $
\mathbb G_{p}(\mathcal A_M,\mathcal{XOR}_k).$
\end{proposition}

\begin{proof}
A clause
\[
\bigl((q_1,\ldots,q_k),(-1)^\beta\bigr)\in M
\]
corresponds exactly to the tuple
\[
(q_1,\ldots,q_k)\in R_\beta^{\mathcal A_M}.
\]
In the homomorphism game this sends $q_i$ to player
$p(q_i)=i$. If the players return $b_1,\ldots,b_k\in\{0,1\}$,
they win precisely when
\[
(b_1,\ldots,b_k)\in R_\beta^{\mathcal{XOR}_k},
\]
that is, precisely when
\[
b_1\oplus\cdots\oplus b_k=\beta.
\]
Equivalently,
\[
\prod_{i=1}^k(-1)^{b_i}=(-1)^\beta,
\]
which is exactly the winning condition of the original $k$-XOR
game.
\end{proof}

\section{Classical strategies}
\label{sec:classical}

We explore the power of classical strategies in $\mathbb{G}_f(\As, \Bs)$.

\begin{definition}
A \textit{deterministic classical strategy} $S_c = \vec{f} = (f_1, \ldots, f_k)$ is defined by functions $f_i: X_i \rightarrow Y_i$ which map inputs to outputs for each player. Hence we have $p(\vec{f}(\vec{x}) | \vec{x}) = 1$.
\end{definition}

\begin{remark}
We could consider probabilistic classical strategies where players are allowed to use shared randomness. However, a simple convexity argument shows that whenever a probabilistic perfect strategy exists a deterministic perfect strategy must also exist.
\end{remark}

Our next result proves a correspondence between classical strategies and homomorphisms. 

\begin{proposition}
\label{prop:classical-strat}
        $\As \rightarrow \Bs$ if and only if $\mathbb{G}_f(\As, \Bs)$ has a perfect classical strategy.
\end{proposition}

\begin{proof}
Let $h:\mathcal{A}\to\mathcal{B}$ be a homomorphism. Define a classical strategy in which every player, on receiving any question $a\in A$, responds with $h(a)$. For the chosen tuple $\vec a\in R^{\mathcal A}$, this strategy produces $\vec{b} = \vec{h}(\vec{a})$ which is in $R^{\mathcal B}$ because $h$ is a homomorphism.

Conversely, suppose that $\omega^c(\mathbb G_f(\mathcal A,\mathcal B))=1$. Then there is a perfect classical strategy. For each player $\alpha\in \operatorname{im}(f)$, let $s_\alpha : f^{-1}(\alpha)\to B$ be the deterministic response function of player $\alpha$. Define a map $h:A\to B$ by 
$h(a):=s_{f(a)}(a).$
This is a well-defined function because $f$ is a function, so each $a\in A$ belongs to exactly one player $f(a)$.
It remains to show that $h$ is a homomorphism. For any $R \in \sigma$ the verifier has a non-zero chance of choosing any tuple $\vec{a} \in R^\As$. Since the strategy is perfect $\vec{h}(\vec{a})\in R^{\mathcal B}$ as required.

\end{proof}

\subsection{Composition and $k$-partite structures}

We now wish to describe the composition of classical strategies. 

Consider a verifier interacting with $k$ spatially separated players. Suppose the players possess perfect strategies for the games $\mathbb{G}_p(\As,\Bs)$ and $\mathbb{G}_q(\Bs,\Cs)$, and wish to combine them into a strategy for a game involving $\As$ and $\Cs$. Since the colourings are fixed, Proposition~\ref{prop:classical-strat} identifies these strategies with homomorphisms $h_1:\As\rightarrow\Bs$ and $h_2:\Bs\rightarrow\Cs.$
If communication between the players were permitted, composition would be immediate. Upon receiving a question $a\in A$, player $p(a)$ would compute the intermediate answer $h_1(a)$ and communicate it to the player responsible for that element in the second game, namely player $q(h_1(a))$. That player would then compute the final answer $h_2(h_1(a))$, yielding the composite homomorphism $h_2\circ h_1:\As\rightarrow\Cs.$ In a non-local game, however, communication is forbidden once play has begun. Consequently, the player $q(h_1(a))$ must be the same player that was responsible for $a$ in the first game. Equivalently, composition is possible only when $q(h_1(a))=p(a)$ for every $a\in A$. This compatibility condition motivates the following definition.

\begin{definition}
The category of $k$-partite structures $\ksig$ has objects which are pairs $(\As, p: G(\As) \rightarrow K_k)$. Morphisms $h: (\As, p) \rightarrow (\Bs, q)$ are homomorphisms $h: \As \rightarrow \Bs$ for which $q \circ h = p$. Identities and composition are defined as in $\Rsig$.
\end{definition}

\begin{proposition}
    $\ksig$ satisfies the axioms of a category. 
\end{proposition}

\begin{proof}
For every object $(\As,p)$, the identity on $\As$ satisfies
$p\circ\id_{\As}=p$.

If $h:(\As,p)\to(\Bs,q)$ and
$g:(\Bs,q)\to(\Cs,r)$ are morphisms, then
\[
r\circ(g\circ h)
=(r\circ g)\circ h
=q\circ h
=p.
\]
Thus morphisms are closed under composition.
The remaining category axioms follow from those of $\Rsig$.
\end{proof}

\subsection{The cofree colouring}
There is a forgetful functor $U: \ksig \rightarrow \Rsig$ which forgets colourings; on objects $U(\As,p)=\As$ and on morphisms $U(h)=h$. We now show that $U$ has a right adjoint, given by a \emph{cofree colouring} construction. This is the universal way of equipping a structure with a colouring so that colour--preserving maps into it are exactly ordinary homomorphisms. 

% Operationally, the cofree colouring records, alongside each potential answer, the identity of the player to whom that answer is addressed.

\begin{definition}
\label{def:cofree}
Let $\As$ be a finite $\sigma$--structure. The \emph{cofree $k$--colouring} $\As^{\cfk}$ is the $\sigma$--structure with universe $A\times [k]$ and, for each $R\in\sigma$ of arity $n$,
\[
\bigl((a_1,\alpha_1),\dots,(a_n,\alpha_n)\bigr)\in R^{\As^{\cfk}}
\quad\Longleftrightarrow\quad
(a_1,\dots,a_n)\in R^{\As}
\ \text{ and }\
\forall i,j\ \bigl(\alpha_i=\alpha_j\Rightarrow a_i=a_j\bigr).
\]
We write $\pi_1 \colon A\times[k]\to\As$ and $\pi_2\colon A\times[k]\to[k]$ for the two projections.
\end{definition}

\begin{lemma}
\label{lem:cofree-welldef}
The assignment $\As\mapsto(\As^{\cfk}, \pi_2)$ extends to a functor $(-)^{\cfk}\colon\Rsig\to\ksig$ by setting, for $g\colon\As\to\Bs$,
$g^{\cfk}(a,\alpha)=(g(a),\alpha).$
\end{lemma}

\begin{proof}
\emph{$\pi_2$ is a colouring.} We must check $\pi_2$ is a graph homomorphism $G(\As^{\cfk})\to K_k$. Suppose $(a,\alpha)\sim (a',\alpha')$, and suppose for contradiction that $\pi_2(a,\alpha)=\pi_2(a',\alpha')$, i.e.\ $\alpha=\alpha'$. The defining condition of $R^{\As^{\cfk}}$ applied to the two coordinates occupied by these vertices forces $a=a'$, whence $(a,\alpha)=(a',\alpha')$, a contradiction as $G(\As^{\cfk})$ is loopless. 

% \emph{$\pi_1$ is a homomorphism.} If $\bigl((b_1,\alpha_1),\dots,(b_r,\alpha_r)\bigr)\in R^{\Bs^{\cfk}}$ then $(b_1,\dots,b_r)\in R^{\Bs}$ by the first clause of the defining condition, and $\pi_1$ returns exactly this tuple.

\emph{Functoriality.} Fix $g\colon\As\to\Bs$. First, $g^{\cfk}$ is a homomorphism: given $(\vec{a}, \vec{\alpha}) \in R^{\As^{\cfk}}$ we have $\vec{a}\in R^{\As}$, so $\vec{g}(\vec{a})\in R^{\Bs}$ as $g$ is a homomorphism; and if $\alpha_i=\alpha_j$ then $a_i=a_j$, hence $g(a_i)=g(a_j)$. Second, $g^{\cfk}$ is colour--preserving: $\pi_2\bigl(g^{\cfk}(a,\alpha)\bigr)=\pi_2(g(a),\alpha)=\alpha=\pi_2(a,\alpha)$. Thus $g^{\cfk}$ is a morphism of $\ksig$. Finally $\id_\Bs^{\cfk}=\id_{\Bs^{\cfk}}$ and $(g'\circ g)^{\cfk}=g'^{\cfk}\circ g^{\cfk}$ hold since both act as $g$ (resp.\ $g'\circ g$) on the first coordinate and as the identity on the second.
\end{proof}

\begin{theorem}\label{thm:cofree-adjunction}
$U\dashv(-)^{\cfk}$.
\end{theorem}

\begin{proof}
We show that $\varepsilon_\Bs:=\pi_1\colon U\bigl(\Bs^{\cfk},\pi_2\bigr)\to\Bs$ is a universal morphism from $U$ to $\Bs$.

First, $\varepsilon_\Bs$ is a homomorphism: if $(\vec{b},\vec{\alpha})\in R^{\Bs^{\cfk}}$, then by the first clause of Definition~\ref{def:cofree} we have $\vec{b}\in R^{\Bs}$.

Now let $(\As,p)\in\ksig$ and let $g\colon U(\As,p)=\As\to\Bs$ be a morphism in $\Rsig$. Since $B\times[k]$ is the product of $B$ and $[k]$ in $\catSet$, there is a unique function $h:=\langle g,p\rangle \colon A\to B\times[k]$ such that
$$
\pi_1\circ h=g
\qquad\text{and}\qquad
\pi_2\circ h=p.
$$
It therefore remains only to check that $h$ is a morphism $(\As,p)\to(\Bs^{\cfk},\pi_2)$ in $\ksig$.

\emph{Colour preservation.} This is immediate from $\pi_2\circ h=p$.

\emph{Relation preservation.} Take $\vec a\in R^{\As}$. We must verify that $(\vec{h}(\vec{a})) = (\vec{g}(\vec{a}), \vec{p}(\vec{a})) \in R^{\Bs^{\cfk}}$. Since $g$ is a homomorphism, we have $\vec g(\vec a)\in R^{\Bs}.$
It remains to verify the second condition in Definition~\ref{def:cofree}. Suppose $p(a_i)=p(a_j)$. If $a_i\neq a_j$, $a_i$ and $a_j$ are adjacent in $G(\As)$ because they appear together in the tuple $\vec{a}$. Since $p$ is a proper colouring, this is impossible. Hence $a_i=a_j$, and therefore $g(a_i)=g(a_j)$ as required. 
\end{proof}

\section{Quantum strategies}
\label{sec:quantum}

We now explore quantum strategies. Quantum theory allows for several alternative mathematical models of such strategies, which are known to give rise to genuinely different classes of correlations. In this work we focus on two of the most widely studied models: the commuting operator model, and the tensor product model.

\begin{definition}

\begin{enumerate}
    \item A \textit{commuting strategy} $S_{co} = (\psi, \vec{P} = (P_1, \ldots, P_k))$ consists of a state $\psi \in \mathbf{H}$ for some potentially infinite dimensional Hilbert space $\mathbf{H}$ and sets of PVMs $P_i(x_i) \in \PVM(\mathbf{H}, Y_i)$ representing players' measurements. In these strategies all measurements of player $i$ must commute with all measurements of player $j$. Then we have $
	p(\vec{y} | \vec{x}) = \psi^\dagger (\Pi_{i\in[k]} P_i(x_i)(y_i)) \psi.$
    \item A \textit{tensor strategy} $S_{t}= (\psi, \vec{P} = (P_1, \ldots, P_k))$ consists of a shared state $\psi \in \vec{\mathbf{H}} = \bigotimes_{i\in [k]} \mathbf{H_i}$ for complex Hilbert spaces and sets $P_i = \{P_i(x_i)\}_{x_i\in X_i}$ where each $P_i(x_i) \in \PVM(\mathbf{H_i}, Y_i)$. Hence we have $p(\vec{y} | \vec{x}) = \psi^\dagger(\bigotimes_{i\in[k]} P_i(x_i)(y_i)) \psi$.
\end{enumerate}
\end{definition}

\begin{remark}
    In general one should allow for strategies which use POVMs instead of PVMs. However, it is known that this choice does not affect the set of strategies \cite{harris2025games}.
\end{remark}

Moving forward we use the term $u$-strategy for $u \in\{c, t, co\}$ to refer to the type of strategy being used, $c$ for classical, $t$ for tensor, and $co$ for commuting. Any $c$-strategy is a $t$-strategy since we can simply ignore entanglement. Moreover, any $t$-strategy is a $co$-strategy. To see this define for each $P_i(x_i) \in \PVM(\mathbf{H_i}, Y_i)$ a canonical embedding $\widehat{P}_i(x_i) \in \PVM(\vec{\mathbf{H}}, Y_i)$ given by
  $$
    \widehat{P}_i(x_i)(y_i)
    :=
    I_{\mathbf{H_1}}\otimes\cdots\otimes
    I_{\mathbf{H_{i-1}}}
    \otimes
    P(x_i)(y_i)
    \otimes
    I_{\mathbf{H_{i+1}}}
    \otimes\cdots\otimes
    I_{\mathbf{H_k}},
  $$
and note that 	$\psi^\dagger (\Pi_{i\in[k]} \widehat{P}_i(x_i)(y_i)) \psi = \psi^\dagger(\bigotimes_{i\in[k]} P_i(x_i)(y_i)) \psi.$

A non-trivial result of Tsirelson shows that if we restrict to finite dimensions $t$-strategies and $co$-strategies coincide \cite{tsirelson2006bell, scholz2008tsirelson}. The breakthrough result \cite{ji2020mip} shows that $co$-strategies form a strictly larger set than $t$-strategies.

\subsection{Commuting operator strategies}

We now characterise $co$-strategies. We begin by defining a notion of state-dependent quantum homomorphism between $k$-partite structures. 

\begin{definition}\label{def:cograde}
A \emph{$k$-partite $C^*$-probability space} is a tuple $\mathfrak{C}=(\mathbf{H};\mathfrak{C}_1,\dots,\mathfrak{C}_k;\psi)$ with $\mathbf{H}$ a Hilbert space, $\mathfrak{C}_1,\dots,\mathfrak{C}_k\subseteq B(\mathbf{H})$ unital $C^*$-subalgebras that pairwise commute, and $\psi\in \mathbf{H}$ a unit vector.
\end{definition}

\begin{definition}
\label{def:sd-qhom}
Let $(\As,p), (\Bs, q)\in\ksig$, and let $\mathfrak{C}$ be a $k$-partite $C^*$-probability space. A \emph{state-dependent quantum commuting homomorphism} from $(\As,p)$ to $(\Bs, q)$, denoted by $(A, p) \rightarrow Q^{co}_{\mathfrak{C}} (B, q)$ is given by a projective measurement $ P(a) \in\ \PVM\!\left(\mathfrak{C}_{p(a)},\,q^{-1}(p(a))\right)$,
for each $a \in A$, subject to the \emph{relation preservation} condition
  \[
    {P}(\vec{a})(\vec{b})\psi = {P}(a_1)(b_1){P}(a_2)(b_2)\cdots {P}(a_n)(b_n)\,\psi \;=\; 0,
  \]

  for every $R\in\sigma$ of arity $n$, every tuple
  $\vec a \in R^{\As}$, and every tuple $\vec b \notin R^{\Bs}$. Here we have set $P(a)(b) = 0 $ whenever $b \notin q^{-1}(p(a))$.

\end{definition}

Note that in the above definition for every $\vec a=(a_1,\ldots,a_n)\in R^\As$, the projective measurements $P(a_1),\ldots,P(a_n)$ pairwise commute. Hence, by Lemma~\ref{lemma:PVM-commeasurability}, $P(\vec a)$ is their joint projective measurement, with $P(\vec a)(\vec b)
=
P(a_1)(b_1)\cdots P(a_n)(b_n).$

We can characterise perfect $co$-strategies in the $\mathbb{G}_p(\As, \Bs)$ game by combining the above definition with the cofree construction from the previous section. 

\begin{theorem}
$(A, p) \rightarrow Q^{co}_{\mathfrak{C}} (B^{\cfk}, \pi_2)$ if and only if  $\mathbb{G}_p(\As, \Bs)$ has a perfect $co$-strategy.
\end{theorem}

\begin{proof}
\noindent\((2)\Rightarrow(1)\).
Suppose first that there exists a $k$-partite $C^*$-probability space $\mathfrak{C}$ and projective measurements
\[
P(a)
\in \PVM\!\bigl(\mathfrak{C}_{p(a)},\,\pi_2^{-1}(p(a))\bigr)
\]
satisfying the relation preservation condition from Definition~\ref{def:sd-qhom}. Since $\pi_2^{-1}(\alpha) = B \times \{\alpha\}$ we may identify each $P(a)$ with a PVM with outcomes in all of $B$. 

Consider the strategy in which, on input $a$, player $p(a)$ performs the measurement $P(a)$ on $\mathfrak{C_{p(a)}}$ and returns the outcome $b \in B$.

Since $p$ is a proper colouring, the measurements associated with the entries of $\vec a$ pairwise commute: distinct entries lie in distinct commuting local algebras, while repeated entries correspond to the same PVM. Hence they admit the joint PVM
\[
P(\vec a)(\vec b)
:=
\prod_{i=1}^n P(a_i)(b_i).
\]

By the Born rule, when the verifier selects the constraint $(R,\vec a)$ the probability that the players answer $\vec b$ is
\[
p(\vec b\mid\vec a)
=
\psi^\dagger{P}(\vec a)(\vec b)\psi.
\]

Since we are dealing with perfect strategies it suffices to verify that $p(\vec b\mid\vec a) = 0$ when $V(\vec{a}, \vec{b}) = 0$. This holds because whenever $\vec b \notin R^\Bs$ the relation preservation condition guarantees that 
\[
{P}({\vec a})({\vec b})\psi=0.
\]
\noindent\((1)\Rightarrow(2)\).
Conversely, suppose there exists a perfect $co$-strategy for $\mathbb G_p(\As,\Bs)$. Thus, there exists a $k$-partite ${C^*}$-probability space $\mathfrak{C}$ and projective measurements $P(a) \in \PVM(\mathbf{\mathfrak{C}_{p(a)}}, B)$ for each $a\in A$. By identifying $B$ with $\pi_2^{-1}(p(a))$ we obtain a candidate map $(\As, p) \rightarrow Q_{\mathfrak{C}} (\Bs^{\cfk}, \pi_2)$ from this data. It remains to check that this candidate satisfies relation preservation. 

Any constraint $(R, \vec{a})$ has a non-zero probability of being chosen by verifier. As above, the colouring $p$ ensures that the operators
\[
P({a_1})(b_1),
\dots,
P({a_n})(b_n),
\]
commute. Since the strategy is perfect, every losing answer $\vec b\notin R^{\Bs}$ occurs with probability zero. Hence
\[
0 = p(\vec{b} \mid \vec{a}) = \psi^\dagger{P}(\vec a)(\vec b)\psi = \psi^\dagger({P}(\vec a)(\vec b))^\dagger({P}(\vec a)(\vec b))\psi = (({P}(\vec a)(\vec b))\psi)^\dagger(({P}(\vec a)(\vec b))\psi) = \|{P}^(\vec a)(\vec b)\psi\|^2,
\]
Therefore ${P}(\vec a)({\vec b})\psi=0$ as required.
\end{proof}

\subsection{The $k$-partite quantum monad}
\label{subsec:kpartite-monad}

In this section we construct a family of endofunctors $Q^{co}_{\mathfrak{C}}$ that can be equipped with the structure of a graded monad. We refer to this as the (state-dependent) $k$-partite quantum commuting monad. Just as classical strategies can be composed in $\ksig$, $co$-strategies can be composed in the Kleisli category of this monad. 

Let $M^{co}_k$ be the class of all $k$-partite $C^*$-probability spaces. Up to strictification these form a monoid with products given by $\mathfrak{C} \otimes \mathfrak{C'} = (\mathbf{H}\otimes \mathbf{H'};\,\mathfrak{C}_\alpha\otimes_{min} \mathfrak{C}'_\alpha;\,\psi\otimes\psi')$ and unit equal to $(\C;\C,\dots,\C;1)$.

\begin{definition}
\label{def:quantum-commuting-monad}
The $k$-partite quantum commuting monad is a $\Sysco$ graded monad defined by the following data:

\begin{enumerate}
    \item For every $\mathfrak{C} \in\Sysco$, there exists an endofunctor $Q^{co}_{\mathfrak{C}}:\cat{R}^k(\sigma)\to\cat{R}^k(\sigma)$.
    \begin{itemize}
        \item The universe of $Q^{co}_{\mathfrak{C}}(\Bs,q)$ is $\coprod_{\alpha\in[k]} \PVM(\mathfrak{C}_\alpha,q^{-1}(\alpha)).$
        \item The colouring sends a PVM in the summand $\PVM(\mathfrak{C}_\alpha,q^{-1}(\alpha))$ to the colour $\alpha$.
        \item A tuple $((\alpha_1, P_1),\ldots,(\alpha_n, P_n))$ belongs to $R^{Q^{co}_{\mathfrak{C}}(\Bs)}$ iff

        \begin{enumerate}
        \item $\forall i\neq j: \alpha_i = \alpha_j \implies P_i = P_j$;
        
        \item for every $\vec b \notin R^\Bs: P_1(b_1)\cdots P_n(b_n)\psi=0.$
        \end{enumerate}
        
        \item For a morphism $h:(\Bs,q)\to(\Cs,r)$ define $Q^{co}_{\mathfrak{C}}(h)(\alpha, P) \in \PVM(\mathfrak{C_\alpha}, r^{-1}(\alpha))$ by
        \[
        Q^{co}_{\mathfrak C}(h)(\alpha, P)(c) = 
        \sum_{b\in h^{-1}(c)}
        P(b).
        \]
\end{itemize}

\item For every $\mathfrak{C}$ and $\mathfrak{C'}$ in $\Sysco$ the multiplication $\mu_{\mathfrak{C, C'}}:Q^{co}_{\mathfrak{C}} Q^{co}_{\mathfrak{C'}}\Rightarrow Q^{co}_{\mathfrak{C \otimes C'}}$ defined by
\[
\mu(\alpha, \Pi)(b):=\sum_{P \in \PVM(\mathfrak{C'_\alpha}, q^{-1}(\alpha))}\Pi(P)\otimes P(b).
\]

\item The unit $\eta: \id_{\ksig} \rightarrow Q^{co}_\mathbf{1} $ is defined by $ \eta(b):=(q(b),\delta_b)$ where $\delta_b\in\PVM(\mathbb C,q^{-1}(q(b)))$ is the Dirac PVM given by $
\delta_b(c)=
\begin{cases}
1&c=b\\
0&c\neq b.
\end{cases}$

\end{enumerate}

\end{definition}

\begin{theorem}
Definition \ref{def:quantum-commuting-monad} satisfies the axioms of a graded monad.    
\end{theorem}

\begin{proof}
Fix $\mathfrak C=(\mathbf H;\mathfrak C_1,\ldots,\mathfrak C_k;\psi)\in\Sysco$. We use the following
conventions throughout this proof.
\begin{itemize}
    \item We write
    $[o=\tilde o]\in\{0,1\}$ for the Kronecker delta, so that
    $P(o)P(\tilde o)=[o=\tilde o]\,P(o)$.
    \item An element of $Q^{co}_{\mathfrak C}(\Bs,q)$ is a pair $(\alpha,P)$ with
    $P\in\PVM(\mathfrak C_\alpha,q^{-1}(\alpha))$, and its colour is $\widehat q(\alpha,P):=\alpha$.
    We identify the elements of colour $\alpha$ with $\PVM(\mathfrak C_\alpha,q^{-1}(\alpha))$. In
    particular, for $\mathfrak C'\in\Sysco$ and an element $(\alpha,\Pi)$ of
    $Q^{co}_{\mathfrak C}Q^{co}_{\mathfrak C'}(\Bs,q)$, we write $\Pi(P)$ for $\Pi(\alpha,P)$.
    \item For a function $f:O\to O'$ and $P\in\PVM(\mathfrak D,O)$, the \emph{pushforward} of $P$
    along $f$ is given by
    \[
    f_*P(o'):=\sum_{o\in f^{-1}(o')}P(o)\qquad(o'\in O'),
    \]
    where $f_*P(o')$ abbreviates $(f_*P)(o')$. For a morphism $h:(\Bs,q)\to(\Cs,r)$ we have
    $h(q^{-1}(\alpha))\subseteq r^{-1}(\alpha)$, since $r\circ h=q$. For
    $P\in\PVM(\mathfrak C_\alpha,q^{-1}(\alpha))$ we write $h_*P$ for the pushforward of $P$ along
    the restriction $q^{-1}(\alpha)\to r^{-1}(\alpha)$ of $h$.
\end{itemize}
In this notation, Definition~\ref{def:quantum-commuting-monad} says that $Q^{co}_{\mathfrak C}(h)$
sends $(\alpha,P)$ to $(\alpha,h_*P)$, and that $\mu_{\mathfrak C,\mathfrak C'}$ sends
$(\alpha,\Pi)$ to $(\alpha,\mu(\alpha,\Pi))$.

\medskip
\noindent\textbf{Step 1: each $Q^{co}_{\mathfrak C}$ is a functor $\ksig\to\ksig$.}

\emph{Pushforwards.}
Let $f:O\to O'$, and let $P\in\PVM(\mathfrak D,O)$ with $\mathfrak D\subseteq B(\mathbf G)$. We check
that $f_*P\in\PVM(\mathfrak D,O')$:
\begin{itemize}
    \item $f_*P$ is finitely supported, since its support is contained in the image under $f$ of
    the support of $P$;
    \item each sum $f_*P(o')$ has finitely many non-zero terms, each a self-adjoint element of
    $\mathfrak D$, so $f_*P(o')$ is a self-adjoint element of $\mathfrak D$;
    \item for $o',o''\in O'$, since distinct fibres of $f$ are disjoint,
    \[
    f_*P(o')\,f_*P(o'')
    =\sum_{o\in f^{-1}(o')}\ \sum_{\tilde o\in f^{-1}(o'')}[o=\tilde o]\,P(o)
    =[o'=o'']\,f_*P(o'),
    \]
    so each $f_*P(o')$ is a projection, and distinct outcomes give orthogonal projections;
    \item $\sum_{o'\in O'}f_*P(o')=\sum_{o\in O}P(o)=I_{\mathbf G}$, since the fibres of $f$
    partition $O$.
\end{itemize}
Moreover, $(\id_O)_*P=P$, and $(g\circ f)_*P=g_*(f_*P)$ for every $g:O'\to O''$, because
$(g\circ f)^{-1}(o'')=\bigsqcup_{o'\in g^{-1}(o'')}f^{-1}(o')$. Finally, let $\mathbf K$ be a
Hilbert space and let $(Y_{o'})_{o'\in O'}$ be a family in $B(\mathbf K)$. Each $o\in O$ lies in
exactly one fibre of $f$, namely $f^{-1}(f(o))$, so regrouping the sum by fibres gives
\begin{equation}
\label{eq:pushforward-regroup}
\sum_{o'\in O'}f_*P(o')\otimes Y_{o'}=\sum_{o\in O}P(o)\otimes Y_{f(o)} .
\end{equation}

\emph{$Q^{co}_{\mathfrak C}(\Bs,q)$ is an object of $\ksig$.}
We must check that $\widehat q$ is a proper colouring of the Gaifman graph of
$Q^{co}_{\mathfrak C}(\Bs,q)$. Suppose that two distinct elements $(\alpha,P)$ and $(\beta,P')$
occur together in a tuple of some relation of $Q^{co}_{\mathfrak C}(\Bs,q)$. If $\alpha=\beta$,
then condition~(a) of Definition~\ref{def:quantum-commuting-monad} gives $P=P'$, and hence
$(\alpha,P)=(\beta,P')$, contradicting distinctness. Therefore $\alpha\neq\beta$.

\emph{$Q^{co}_{\mathfrak C}(h)$ is a morphism of $\ksig$.}
Fix a morphism $h:(\Bs,q)\to(\Cs,r)$. By the above,
$h_*P\in\PVM(\mathfrak C_\alpha,r^{-1}(\alpha))$ for every element $(\alpha,P)$ of
$Q^{co}_{\mathfrak C}(\Bs,q)$, so $Q^{co}_{\mathfrak C}(h)$ is a well-defined, colour-preserving
map. We record that
\[
h_*P(c)=\sum_{b\in h^{-1}(c)}P(b)\qquad\text{for every }c\in C.
\]
Indeed, if $c\in r^{-1}(\alpha)$, then $h^{-1}(c)\subseteq q^{-1}(\alpha)$; and if
$c\notin r^{-1}(\alpha)$, then every $b\in h^{-1}(c)$ satisfies $q(b)=r(c)\neq\alpha$, so both
sides are $0$.

Now let $R\in\sigma$ have arity $n$, and let
$\bigl((\alpha_1,P_1),\ldots,(\alpha_n,P_n)\bigr)\in R^{Q^{co}_{\mathfrak C}(\Bs,q)}$. We verify
conditions~(a) and~(b) for the image tuple
$\bigl((\alpha_1,h_*P_1),\ldots,(\alpha_n,h_*P_n)\bigr)$. For~(a), if $\alpha_i=\alpha_j$, then
$P_i=P_j$ by condition~(a) for the original tuple, and hence $h_*P_i=h_*P_j$. For~(b), let
$\vec c\notin R^{\Cs}$. Expanding the product gives
\[
h_*P_1(c_1)\cdots h_*P_n(c_n)\,\psi
=\sum_{\substack{b_i\in h^{-1}(c_i)\\ i\in[n]}}P_1(b_1)\cdots P_n(b_n)\,\psi .
\]
Each summand has $\vec h(\vec b)=\vec c\notin R^{\Cs}$, so $\vec b\notin R^{\Bs}$ because $h$ is a
homomorphism. Hence every summand vanishes by condition~(b) for the original tuple.

\emph{Functoriality.}
The restriction of $\id_{(\Bs,q)}$ to $q^{-1}(\alpha)$ is the identity, and the restriction of a
composite $(\Bs,q)\xrightarrow{h}(\Cs,r)\xrightarrow{g}(\mathcal D,t)$ is the composite of the
restrictions. The properties of pushforwards established above therefore give
$Q^{co}_{\mathfrak C}(\id_{(\Bs,q)})=\id_{Q^{co}_{\mathfrak C}(\Bs,q)}$ and
$Q^{co}_{\mathfrak C}(g\circ h)=Q^{co}_{\mathfrak C}(g)\circ Q^{co}_{\mathfrak C}(h)$.

\medskip
\noindent\textbf{Step 2: the unit $\eta:\id_{\ksig}\Rightarrow Q^{co}_{\mathbf 1}$.}

The only projections in $\C=B(\C)$ are $0$ and $1$. A finitely supported PVM with values in
$\{0,1\}$ and total sum $1$ takes the value $1$ at exactly one outcome, so the elements of
$\PVM(\C,q^{-1}(\alpha))$ are exactly the Dirac PVMs $\delta_b$ with $b\in q^{-1}(\alpha)$. In
particular, $\eta_{(\Bs,q)}(b)=(q(b),\delta_b)$ is an element of $Q^{co}_{\mathbf 1}(\Bs,q)$ of
colour $q(b)$, so $\eta_{(\Bs,q)}$ is colour-preserving.

\emph{$\eta_{(\Bs,q)}$ preserves relations.}
Let $R\in\sigma$ have arity $n$, and let $\vec b\in R^{\Bs}$. For~(a), suppose $q(b_i)=q(b_j)$. If
$b_i\neq b_j$, then $b_i$ and $b_j$ are adjacent in $G(\Bs)$, contradicting that $q$ is a proper
colouring. Hence $b_i=b_j$, and so $\delta_{b_i}=\delta_{b_j}$. For~(b), let
$\vec c\notin R^{\Bs}$. Since $\vec c\neq\vec b$ and the state of $\mathbf 1$ is $1\in\C$,
\[
\delta_{b_1}(c_1)\cdots\delta_{b_n}(c_n)\cdot 1=[b_1=c_1]\cdots[b_n=c_n]=0 .
\]

\emph{Naturality.}
Let $h:(\Bs,q)\to(\Cs,r)$ be a morphism, and let $b\in B$. For every $c\in C$,
\[
h_*\delta_b(c)=\sum_{b'\in h^{-1}(c)}\delta_b(b')=[h(b)=c]=\delta_{h(b)}(c).
\]
Since $r(h(b))=q(b)$, it follows that
\[
Q^{co}_{\mathbf 1}(h)\bigl(\eta_{(\Bs,q)}(b)\bigr)=(q(b),\delta_{h(b)})=\eta_{(\Cs,r)}(h(b)).
\]

\medskip
\noindent\textbf{Step 3: the multiplication $\mu_{\mathfrak C,\mathfrak C'}:Q^{co}_{\mathfrak C}Q^{co}_{\mathfrak C'}\Rightarrow Q^{co}_{\mathfrak C\otimes\mathfrak C'}$.}

\emph{$\mu(\alpha,\Pi)$ is a PVM.}
Let $(\alpha,\Pi)$ be an element of $Q^{co}_{\mathfrak C}Q^{co}_{\mathfrak C'}(\Bs,q)$, so that
\[
\Pi\in\PVM\bigl(\mathfrak C_\alpha,\PVM(\mathfrak C'_\alpha,q^{-1}(\alpha))\bigr).
\]
For each $b\in B$, the sum
\[
\mu(\alpha,\Pi)(b)=\sum_{P}\Pi(P)\otimes P(b)
\]
has finitely many non-zero terms, each of which is a self-adjoint element of
$\mathfrak C_\alpha\otimes_{\min}\mathfrak C'_\alpha$. For $b,b'\in q^{-1}(\alpha)$,
\[
\begin{aligned}
\mu(\alpha,\Pi)(b)\,\mu(\alpha,\Pi)(b')
&=\sum_{P,\tilde P}\Pi(P)\Pi(\tilde P)\otimes P(b)\tilde P(b')\\
&=\sum_{P}\Pi(P)\otimes P(b)P(b')
=[b=b']\,\mu(\alpha,\Pi)(b),
\end{aligned}
\]
and
\[
\sum_{b\in q^{-1}(\alpha)}\mu(\alpha,\Pi)(b)
=\sum_{P}\Pi(P)\otimes I_{\mathbf H'}
=I_{\mathbf H}\otimes I_{\mathbf H'}.
\]
Finally, $\mu(\alpha,\Pi)(b)\neq 0$ only if $P(b)\neq 0$ for one of the finitely many $P$ with
$\Pi(P)\neq 0$, so $\mu(\alpha,\Pi)$ is finitely supported. Hence
$\mu(\alpha,\Pi)\in\PVM\bigl(\mathfrak C_\alpha\otimes_{\min}\mathfrak C'_\alpha,q^{-1}(\alpha)\bigr)$,
and the component of $\mu_{\mathfrak C,\mathfrak C'}$ at $(\Bs,q)$ is a well-defined,
colour-preserving map.

\emph{Preservation of relations.}
Let $R\in\sigma$ have arity $n$, and let
\[
\bigl((\alpha_1,\Pi_1),\ldots,(\alpha_n,\Pi_n)\bigr)\in
R^{Q^{co}_{\mathfrak C}Q^{co}_{\mathfrak C'}(\Bs,q)}.
\]
For~(a), if $\alpha_i=\alpha_j$, then
$\Pi_i=\Pi_j$, and hence $\mu(\alpha_i,\Pi_i)=\mu(\alpha_j,\Pi_j)$. For~(b), let
$\vec c\notin R^{\Bs}$. Since $(X\otimes Y)(X'\otimes Y')=XX'\otimes YY'$ for all bounded
operators, expanding the product gives
\[
\begin{aligned}
&\mu(\alpha_1,\Pi_1)(c_1)\cdots\mu(\alpha_n,\Pi_n)(c_n)\,(\psi\otimes\psi')\\
&\qquad=\sum_{P_1,\ldots,P_n}
\bigl(\Pi_1(P_1)\cdots\Pi_n(P_n)\,\psi\bigr)\otimes\bigl(P_1(c_1)\cdots P_n(c_n)\,\psi'\bigr),
\end{aligned}
\]
where each $P_i$ ranges over $\PVM(\mathfrak C'_{\alpha_i},q^{-1}(\alpha_i))$. Fix a term of this
sum, and consider the tuple $\vec x:=\bigl((\alpha_1,P_1),\ldots,(\alpha_n,P_n)\bigr)$ of elements
of $Q^{co}_{\mathfrak C'}(\Bs,q)$.
\begin{itemize}
    \item If $\vec x\notin R^{Q^{co}_{\mathfrak C'}(\Bs,q)}$, then condition~(b) for the tuple
    $\bigl((\alpha_1,\Pi_1),\ldots,(\alpha_n,\Pi_n)\bigr)$, applied to $\vec x$, gives
    $\Pi_1(P_1)\cdots\Pi_n(P_n)\,\psi=0$.
    \item If $\vec x\in R^{Q^{co}_{\mathfrak C'}(\Bs,q)}$, then, since $\vec c\notin R^{\Bs}$,
    condition~(b) for $\vec x$ gives $P_1(c_1)\cdots P_n(c_n)\,\psi'=0$.
\end{itemize}
In either case the term vanishes, so the whole sum is $0$. Hence the component of
$\mu_{\mathfrak C,\mathfrak C'}$ at $(\Bs,q)$ is a morphism of $\ksig$.

\emph{Naturality.}
Let $h:(\Bs,q)\to(\Cs,r)$ be a morphism, and let $(\alpha,\Pi)$ be an element of
$Q^{co}_{\mathfrak C}Q^{co}_{\mathfrak C'}(\Bs,q)$. On elements of colour $\alpha$, the map
$Q^{co}_{\mathfrak C'}(h)$ is $P\mapsto h_*P$. Hence
$Q^{co}_{\mathfrak C}Q^{co}_{\mathfrak C'}(h)$ sends $(\alpha,\Pi)$ to
$\bigl(\alpha,(h_*)_*\Pi\bigr)$. Let $c\in r^{-1}(\alpha)$. Applying
\eqref{eq:pushforward-regroup} with $f=h_*$ and $Y_{P'}=P'(c)$ gives
\[
\mu\bigl(\alpha,(h_*)_*\Pi\bigr)(c)
=\sum_{P'}\bigl((h_*)_*\Pi\bigr)(P')\otimes P'(c)
=\sum_{P}\Pi(P)\otimes h_*P(c).
\]
On the other hand,
\[
h_*\bigl(\mu(\alpha,\Pi)\bigr)(c)
=\sum_{b\in h^{-1}(c)}\ \sum_{P}\Pi(P)\otimes P(b)
=\sum_{P}\Pi(P)\otimes h_*P(c).
\]
Both sides are elements of colour $\alpha$ with the same PVM, so
\[
(\mu_{\mathfrak C,\mathfrak C'})_{(\Cs,r)}\circ Q^{co}_{\mathfrak C}Q^{co}_{\mathfrak C'}(h)
=Q^{co}_{\mathfrak C\otimes\mathfrak C'}(h)\circ(\mu_{\mathfrak C,\mathfrak C'})_{(\Bs,q)} .
\]

\medskip
\noindent\textbf{Step 4: the unit laws.}
In this step we use the identifications $\C\otimes\mathbf H\cong\mathbf H\cong\mathbf H\otimes\C$,
under which $\mathbf 1\otimes\mathfrak C=\mathfrak C=\mathfrak C\otimes\mathbf 1$. Let $(\alpha,P)$
be an element of $Q^{co}_{\mathfrak C}(\Bs,q)$, and let $b\in q^{-1}(\alpha)$.

\emph{Left unit: $\mu_{\mathbf 1,\mathfrak C}\circ\eta_{Q^{co}_{\mathfrak C}}=\id_{Q^{co}_{\mathfrak C}}$.}
The component of $\eta$ at $Q^{co}_{\mathfrak C}(\Bs,q)$ sends $(\alpha,P)$ to $(\alpha,\delta_P)$,
where $\delta_P\in\PVM\bigl(\C,\PVM(\mathfrak C_\alpha,q^{-1}(\alpha))\bigr)$ is the Dirac PVM
at $P$. Hence
\[
\mu(\alpha,\delta_P)(b)=\sum_{P'}\delta_P(P')\otimes P'(b)=1\otimes P(b)=P(b).
\]

\emph{Right unit: $\mu_{\mathfrak C,\mathbf 1}\circ Q^{co}_{\mathfrak C}\eta=\id_{Q^{co}_{\mathfrak C}}$.}
On elements of colour $\alpha$, the map $\eta_{(\Bs,q)}$ restricts to
$\eta^\alpha:q^{-1}(\alpha)\to\PVM(\C,q^{-1}(\alpha))$, $c\mapsto\delta_c$. Hence
$Q^{co}_{\mathfrak C}(\eta_{(\Bs,q)})$ sends $(\alpha,P)$ to $(\alpha,\eta^\alpha_*P)$. Applying
\eqref{eq:pushforward-regroup} with $f=\eta^\alpha$ and $Y_D=D(b)$, where $D$ ranges over
$\PVM(\C,q^{-1}(\alpha))$, gives
\[
\mu(\alpha,\eta^\alpha_*P)(b)
=\sum_{D}\eta^\alpha_*P(D)\otimes D(b)
=\sum_{c\in q^{-1}(\alpha)}P(c)\otimes\delta_c(b)
=P(b)\otimes 1
=P(b).
\]
Both unit laws therefore hold.

\medskip
\noindent\textbf{Step 5: associativity.}
Let $\mathfrak C',\mathfrak C''\in\Sysco$, with underlying Hilbert spaces $\mathbf H'$ and
$\mathbf H''$, and let $(\alpha,\Xi)$ be an element of
$Q^{co}_{\mathfrak C}Q^{co}_{\mathfrak C'}Q^{co}_{\mathfrak C''}(\Bs,q)$. Thus
\[
\Xi\in\PVM\Bigl(\mathfrak C_\alpha,\ \PVM\bigl(\mathfrak C'_\alpha,\ \PVM(\mathfrak C''_\alpha,q^{-1}(\alpha))\bigr)\Bigr).
\]
Below, $P$ ranges over $\PVM(\mathfrak C''_\alpha,q^{-1}(\alpha))$, and $\Pi$ ranges over the
PVMs in $\mathfrak C'_\alpha$ with outcomes in this set. We compare the two composites
\[
\mu_{\mathfrak C\otimes\mathfrak C',\mathfrak C''}\circ\mu_{\mathfrak C,\mathfrak C'}Q^{co}_{\mathfrak C''}
\qquad\text{and}\qquad
\mu_{\mathfrak C,\mathfrak C'\otimes\mathfrak C''}\circ Q^{co}_{\mathfrak C}\mu_{\mathfrak C',\mathfrak C''}
\]
at $(\alpha,\Xi)$, by computing their values at each $b\in q^{-1}(\alpha)$.

For the first composite, the component of $\mu_{\mathfrak C,\mathfrak C'}$ at
$Q^{co}_{\mathfrak C''}(\Bs,q)$ sends $(\alpha,\Xi)$ to $(\alpha,\Theta)$, where
$\Theta(P)=\sum_{\Pi}\Xi(\Pi)\otimes\Pi(P)$. Applying
$\mu_{\mathfrak C\otimes\mathfrak C',\mathfrak C''}$ gives
\[
\mu(\alpha,\Theta)(b)
=\sum_{P}\Theta(P)\otimes P(b)
=\sum_{\Pi,P}\Xi(\Pi)\otimes\Pi(P)\otimes P(b).
\]

For the second composite, let $\mu^\alpha$ denote the map $\Pi\mapsto\mu(\alpha,\Pi)$, which is
$\mu_{\mathfrak C',\mathfrak C''}$ on elements of colour $\alpha$. Then
$Q^{co}_{\mathfrak C}(\mu_{\mathfrak C',\mathfrak C''})$ sends $(\alpha,\Xi)$ to
$(\alpha,\mu^\alpha_*\Xi)$. Applying $\mu_{\mathfrak C,\mathfrak C'\otimes\mathfrak C''}$, and then
\eqref{eq:pushforward-regroup} with $f=\mu^\alpha$ and $Y_\Phi=\Phi(b)$, where $\Phi$ ranges over
$\PVM(\mathfrak C'_\alpha\otimes_{\min}\mathfrak C''_\alpha,q^{-1}(\alpha))$, gives
\[
\begin{aligned}
\mu(\alpha,\mu^\alpha_*\Xi)(b)
&=\sum_{\Phi}\mu^\alpha_*\Xi(\Phi)\otimes\Phi(b)
=\sum_{\Pi}\Xi(\Pi)\otimes\mu(\alpha,\Pi)(b)\\
&=\sum_{\Pi,P}\Xi(\Pi)\otimes\Pi(P)\otimes P(b).
\end{aligned}
\]
Under the identification
$(\mathbf H\otimes\mathbf H')\otimes\mathbf H''\cong\mathbf H\otimes(\mathbf H'\otimes\mathbf H'')$,
the two composites agree. This is the associativity axiom.

\end{proof}
\begin{definition}
    The Kleisli category of $Q^{co}_{\mathfrak C}$ denoted by $\Kl(Q^{co}_{\mathfrak{C}})$ is defined as follows\footnote{There is a more involved definition of the Kleisli category of a graded monad available in \cite{fujii2016towards}, but for our purposes this definition will suffice.}:
    \begin{itemize}
        \item Objects are the same objects as $\ksig$;
        \item Morphisms between $(\As, p)$ and $(\Bs, q)$ are morphisms $(\As, p) \rightarrow Q^{co}_{\mathfrak{C}} (\Bs, q)$ in $\ksig$;
        \item The composition of $f:(\As, p) \rightarrow Q^{co}_{\mathfrak{C}} (\Bs, q)$ and $g:(\Bs, q) \rightarrow Q^{co}_{\mathfrak{C'}} (\Cs, r)$ is given by $(\As, p) \overset{f}{\rightarrow} Q^{co}_{\mathfrak{C}} (\Bs, q) \overset{Q^{co}_{\mathfrak{C}}(g)}{\rightarrow}Q^{co}_{\mathfrak{C}} Q^{co}_{\mathfrak{C'}} (\Cs, r) \overset{\mu}{\rightarrow}Q^{co}_{\mathfrak{C}\otimes \mathfrak{C'}}(\Cs, r)$;
        \item The unit is given by $\eta_{(\As, p)}: (\As, p) \rightarrow Q^{co}_{{1}} (\As, p)$.
    \end{itemize}
     
\end{definition}

Under this definition a Kleisli morphism is readily seen to be a state-dependent $k$-partite quantum commuting homomorphism as per definition \ref{def:sd-qhom}.

\subsection{Tensor product strategies}

In order to characterise $t$-strategies we work with $k$-partite Hilbert spaces.

\begin{definition}
    A $k$-partite pointed Hilbert space is a pair $(\mathbf{H}, \psi)$ where $\mathbf{H}$ is a Hilbert space which decomposes as $\mathbf H = \bigotimes_{i \in [k]} \mathbf{H_i}$ and $\psi$ is a pure state on $\mathbf{H}$. 
\end{definition}

Up to strictification $k$-partite pointed Hilbert spaces form a monoid $\Syst$ with product $
(\bigotimes_{i\in[k]}\mathbf{H_i},\psi)\otimes (\bigotimes_{i\in[k]}\mathbf{G_i},\phi)
:= (\bigotimes_{i\in[k]}\mathbf{H_i} \otimes \mathbf{G_i},\psi \otimes \phi)$
and unit grade by $\mathbf 1:=(\mathbb C,\ldots,\mathbb C;1)$. 

% There is a canonical inclusion $
% \iota:\Syst\hookrightarrow\Sysco $
% given by
% \[
% \iota\bigl((\mathbf H_1,\ldots,\mathbf H_k;\psi)\bigr)
% =
% \left(
% \bigotimes_{\alpha=1}^k \mathbf H_\alpha;\
% \mathfrak C_1,\ldots,\mathfrak C_k;\
% \psi
% \right),
% \]
% where
% \[
% \mathfrak C_\alpha
% =
% I_{\mathbf H_1}\otimes\cdots\otimes
% I_{\mathbf H_{\alpha-1}}
% \otimes B(\mathbf H_\alpha)
% \otimes
% I_{\mathbf H_{\alpha+1}}\otimes\cdots\otimes
% I_{\mathbf H_k}.
% \]

\begin{definition}
\label{def:quantum-tensor-monad}
The $k$-partite quantum tensor monad is a $\Syst$ graded monad defined exactly as in definition \ref{def:quantum-commuting-monad}, with each local algebra $\mathfrak C_\alpha$ replaced by $B(\mathbf{H}_i)$.
\end{definition}

We omit the proofs of the following theorems as they are almost verbatim the same as their commuting operator analogues.

\begin{theorem}
    Definition \ref{def:quantum-tensor-monad} satisfies the axioms of a graded monad. 
\end{theorem}

\begin{theorem}
$(A, p) \rightarrow Q^{t}_{\mathbf{H}, \psi} (\Bs^{\cfk}, \pi_2)$ if and only if $\mathbb{G}_p(\As, \Bs)$ has a perfect $t$-strategy.
\end{theorem}

% If we additionally write $Q^c_1 = \id$ to represent the identity monad we obtain the following result, which uniformly characterises classical, tensor, and commuting strategies. 

% \begin{theorem}
%     Take $u \in \{c, t, co\}$. The following are equivalent:
%     \begin{enumerate}
%         \item There exists a perfect $u$-strategy in $\mathbb{G}_p(\As, \Bs)$.
%         \item $(A, p) \rightarrow Q^{u}_{g} (\Bs^{\cfk}, \pi_2).$
%     \end{enumerate}
% \end{theorem}

\section{Quantum advantage beyond $\mathcal{XOR}$}
\label{sec:beyond-xor}

% Our developments thus far allow us to define $k$-partite quantum relaxations of the $\mathsf{CSP}$ and $\mathsf{CSP{(\Bs)}}$ decision problems\footnote{Note that the colouring $p$ is part of the input data.}:

% \begin{description}
%     \item[$\mathsf{k-QuCSP^u}$]: Given as input $(\As, p) \in \ksig$ and $\Bs \in \Rsig$ decide if $(\As, p) \rightarrow Q^u_g (\Bs^{\cfk},\pi_2$).
% \end{description}

% \begin{description}
%     \item[$\mathsf{k-QuCSP^u}$($\Bs$)]: Given as input $(\As, p) \in \ksig$ decide if $(\As, p) \rightarrow Q^u_g(\Bs^{\cfk},\pi_2$).
% \end{description}

Our developments thus far allow us to define $k$-partite quantum advantage for CSPs.

\begin{definition}
\label{def:k-partite-advantage}

For $u \in \{t, co\}$, a CSP instance $(\As, \Bs)$ admits (pseudotelepathic) $k$-partite $u$-advantage if $\As \not \rightarrow \Bs$ and $(\As,p) \rightarrow Q^u_g (\Bs^{\cfk}, \pi_2)$ (for some colouring $p$ and grade $g$). $\mathsf{CSP(\Bs)}$ admits advantage if there exists some $\As$ such that $(\As, \Bs)$ admits advantage. 
\end{definition}

Perhaps surprisingly, there do not appear to be many pseudotelepathic non-local games in the literature which are not instances of synchronous games\footnote{This claim should not be confused with the fact that non-synchronous variants of games such as magic square often appear in the literature. With some abuse of terminology we can still treat these variants as synchronous games, since it can be shown that they admit perfect $u$-strategies iff their synchronous versions do.}, or $k$-XOR games. Thus, one may wonder, are there any CSP templates other than $\mathsf{CSP(\mathcal{XOR}_k)}$ which admit $k$-partite advantage? We answer this question positively by exploring the class of inner-product games recently introduced in \cite{lalonde2026maximall}.

\begin{definition}[Inner-product data]
\label{def:inner-product-data}
An \emph{inner-product datum} is a tuple
\[
\mathbb{D}=(m;U_1,\ldots,U_{n_1};V_1,\ldots,V_{n_2};S),
\]
where $m\geq 2$, the $U_x,V_y\in B(\C^m)$ are unitary, and
$S\in B(\C^m)$ is non-zero and positive semidefinite.
The associated non-local game $G_\mathbb{D}$ has input sets $[n_1]$ and $[n_2]$,
output sets $[m]$ and $[m]$, uniform question distribution, and winning
predicate $ V\bigl((x,y),(a,b)\bigr)=1 \quad\Longleftrightarrow\quad \bra{a}U_x^\dagger S V_y\ket{b}\neq 0.$
\end{definition}

We now translate $G_\mathbb{D}$ into a $2$-homomorphism game. Fix $n_1,n_2\in\N$, and let $\sigma_{n_1,n_2}:=\{R_{x,y}:x\in[n_1],\ y\in[n_2]\}$
be the relational signature consisting of $n_1n_2$ binary relation symbols. We describe the following structures on this signature:
\begin{itemize}
    \item The \emph{bipartite pattern} $\mathcal I_{n_1,n_2}$ has universe $
    I:=\{u_1,\ldots,u_{n_1}\}\sqcup\{v_1,\ldots,v_{n_2}\},$
    and relations $ R_{x,y}^{\mathcal I_{n_1,n_2}}:=\{(u_x,v_y)\}.$ We define a colouring $p:G(I)\to K_2$ by setting $p(u_x)=1$ and $p(v_y)=2.$ This makes $(\mathcal I_{n_1,n_2}, p)$ a $k$-partite structure.

    \item The \emph{inner-product template}
    $\Bs_\mathbb{D}$ is the $\sigma_{n_1,n_2}$-structure with universe $[m]$ and
    \[
    R_{x,y}^{\Bs_\mathbb{D}}
    :=\bigl\{(a,b)\in[m]^2:\bra{a}U_x^\dagger S V_y\ket{b}\neq0\bigr\}
    \]
\end{itemize}

\begin{proposition}
\label{prop:inner-product-game-translation}
The homomorphism game $\mathbb G_p(\mathcal I_{n_1,n_2},\Bs_\mathbb{D})$ is exactly the
inner-product game $G_\mathbb{D}$.
\end{proposition}

\begin{proof}
In $\mathbb G_p(\mathcal I_{n_1,n_2},\Bs_D)$ the verifier chooses a relation
symbol $R_{x,y}$ and a tuple in $R_{x,y}^{\mathcal I_{n_1,n_2}}$ uniformly at
random. Since each such relation is the singleton $\{(u_x,v_y)\}$, this is
precisely a uniform choice of $(x,y)\in[n_1]\times[n_2]$. The questions $u_x$
and $v_y$ are sent to players $p(u_x)=1$ and $p(v_y)=2$, respectively. Each
player answers with an element of $[m]$, and $(a,b)$ is accepted exactly when
$
(a,b)\in R_{x,y}^{\Bs_D}.$
\end{proof}

Consider now the inner-product datum

    $$
    \mathbb{L} = (m = 6; U_1, U_2, U_3, U_4;V_1 , V_2, V_3; S = diag(1,1,2,2,2,2))
    $$
    
    where the unitaries are given by
\[
U_1=
\begin{pmatrix}
0&0&0&0&s&s\\
0&0&0&0&-s&s\\
0&0&0&1&0&0\\
0&0&1&0&0&0\\
0&1&0&0&0&0\\
1&0&0&0&0&0
\end{pmatrix},
\qquad
U_2=
\begin{pmatrix}
s&s&0&0&0&0\\
-s&s&0&0&0&0\\
0&0&\frac12&\frac12&\frac12&\frac12\\
0&0&-\frac12&-\frac12&\frac12&\frac12\\
0&0&-\frac12&\frac12&-\frac12&\frac12\\
0&0&\frac12&-\frac12&-\frac12&\frac12
\end{pmatrix}.
\]
\[
U_3=\frac12
\begin{pmatrix}
0&0&1&1&1&1\\
0&0&-1&-1&1&1\\
1&1&0&0&-1&1\\
-1&-1&0&0&-1&1\\
-1&1&-1&1&0&0\\
-1&1&1&-1&0&0
\end{pmatrix},
\qquad
U_4=\frac12
\begin{pmatrix}
0&0&1&1&1&1\\
0&0&-1&-1&1&1\\
1&1&-1&1&0&0\\
1&1&1&-1&0&0\\
-1&1&0&0&-1&1\\
1&-1&0&0&-1&1
\end{pmatrix}.
\]
\[
V_1=s
\begin{pmatrix}
0&0&0&0&1&1\\
0&0&0&0&-1&1\\
0&0&1&1&0&0\\
0&0&-1&1&0&0\\
1&1&0&0&0&0\\
-1&1&0&0&0&0
\end{pmatrix},
\qquad
V_2=
\begin{pmatrix}
0&0&\frac12&\frac12&\frac12&\frac12\\
0&0&-\frac12&-\frac12&\frac12&\frac12\\
0&0&-\frac12&\frac12&-\frac12&\frac12\\
s&s&0&0&0&0\\
-s&s&0&0&0&0\\
0&0&\frac12&-\frac12&-\frac12&\frac12
\end{pmatrix},
\]
\[
V_3=
\begin{pmatrix}
0&0&\frac12&\frac12&\frac12&\frac12\\
0&0&-\frac12&-\frac12&\frac12&\frac12\\
0&s&-\frac12&\frac12&0&0\\
s&0&0&0&-\frac12&\frac12\\
0&s&\frac12&-\frac12&0&0\\
s&0&0&0&\frac12&-\frac12
\end{pmatrix}
\]
with $s:=\frac{1}{\sqrt2}.$

The main result of \cite{lalonde2026maximall} shows that $G_\mathbb{L}$ has a perfect $t$-strategy but no perfect $t$-strategy involving a maximally entangled state, and no perfect classical strategy. By performing the translation described in the previous proposition to $G_\mathbb{L}$ we obtain the inner-product template $\mathcal{B}_\mathbb{L}$.

To prove our next result we require some concepts from CSP theory.

\begin{definition}
Let $\sigma$ be a relational signature. A \emph{primitive positive (pp)
formula} over $\sigma$ is a first-order formula of the form
\[
\exists y_1\cdots\exists y_m\,\bigl(\alpha_1\wedge\cdots\wedge\alpha_\ell\bigr),
\]
where each $\alpha_i$ is either an atomic formula $R(z_1,\ldots,z_r)$ with
$R\in\sigma$, or an equality $z_1=z_2$. Let $\Bs$ be a $\sigma$-structure. A
relation $S\subseteq B^n$ is \emph{pp-definable} in $\Bs$ if there is a
pp-formula $\varphi(x_1,\ldots,x_n)$ over $\sigma$ such that
\[
S=\varphi^{\Bs}:=\{\vec b\in B^n:\Bs\models\varphi(\vec b)\}.
\]
\end{definition}

\begin{theorem}[\cite{barto2017polymorphisms}]
\label{thm:pp-reduction}
Let $\Bs$ be a finite $\sigma$-structure, and let $\mathcal C$ be a structure
over a finite signature with universe $B$, each of whose relations is
pp-definable in $\Bs$. Then $\mathsf{CSP}(\mathcal C)$ reduces to
$\mathsf{CSP}(\Bs)$ in polynomial time.
\end{theorem}

We can now prove the main theorem of this section.

\begin{theorem}
    $\mathsf{CSP(\mathcal{B}_\mathbb{L})}$ is NP-complete and admits $2$-partite $t$-advantage. 
\end{theorem}

\begin{proof}

\cite{lalonde2026maximall} shows that the game $G_\mathbb{L}$ exhibits pseudotelepathy. Via proposition \ref{prop:inner-product-game-translation} this means that the game $G_p(\mathcal{I}_{4,3}, \Bs_\mathbb{L})$ exhibits pseudotelepathy. Hence CSP($\Bs_\mathbb{L})$ has quantum advantage. 

It remains to prove NP-completeness. Membership in NP is clear, so it suffices
to prove NP-hardness, which we do via a pp-definition.

The NP-complete template we reduce from is
\[
\mathcal N:=\bigl([6];\,N\bigr),
\qquad
N:=\bigl\{(x,y,z)\in\{1,2\}^3 : x,y,z\text{ are not all equal}\bigr\}.
\]
Let $\mathrm{NAE}_3:=\{0,1\}^3\setminus\{(0,0,0),(1,1,1)\}$. The problem
$\mathsf{CSP}(\{0,1\};\mathrm{NAE}_3)$ is NP-complete
\cite{schaefer1978complexity}. Moreover, $\mathcal N$ is homomorphically
equivalent to $(\{0,1\};\mathrm{NAE}_3)$:
\begin{itemize}
    \item the map $0\mapsto 1$, $1\mapsto 2$ is a homomorphism
    $(\{0,1\};\mathrm{NAE}_3)\to\mathcal N$;
    \item the map sending $2\mapsto 1$ and every other element of $[6]$ to
    $0$ is a homomorphism $\mathcal N\to(\{0,1\};\mathrm{NAE}_3)$, since
    $N\subseteq\{1,2\}^3$.
\end{itemize}
Homomorphically equivalent structures have the same CSP, so
$\mathsf{CSP}(\mathcal N)$ is NP-complete.

By Theorem~\ref{thm:pp-reduction}, it now suffices to show that $N$ is
pp-definable in $\Bs_{\mathbb L}$. We use the following notation:
\begin{itemize}
    \item $R_{x,y}:=R^{\Bs_{\mathbb L}}_{x,y}=\supp(U_x^\dagger S V_y)$;
    \item for a binary relation $R$ on $[6]$,
    $R(a):=\{b:(a,b)\in R\}$ and $R^{-1}(b):=\{a:(a,b)\in R\}$.
\end{itemize}
A direct computation gives
\[
\begin{array}{lll}
\{a:(a,a)\in R_{2,1}\}=\{3\}, & R_{1,2}(3)=\{1,2\}, &\\[2pt]
R_{1,3}(1)=\{1,5,6\}, & R^{-1}_{3,2}(1)=\{2,3,4,5,6\}, & R_{3,3}(1)=\{1,3,4\},\\[2pt]
R_{1,3}(2)=\{2,3,4\}, & R^{-1}_{3,2}(2)=\{1,3,4,5,6\}, & R_{3,3}(2)=\{2,5,6\}.
\end{array}
\]

First, let
\[
\delta(x):=\exists z\,\bigl(R_{2,1}(z,z)\wedge R_{1,2}(z,x)\bigr).
\]
Since $3$ is the unique loop of $R_{2,1}$, we have
$\delta^{\Bs_{\mathbb L}}=R_{1,2}(3)=\{1,2\}$.

Next, let
\[
\theta(x,y,z):=\delta(x)\wedge\delta(y)\wedge\delta(z)\wedge
\exists w\,\bigl(R_{1,3}(x,w)\wedge R_{3,2}(w,y)\wedge R_{3,3}(z,w)\bigr).
\]
After renaming bound variables and moving the quantifiers to the front,
$\theta$ is a pp-formula. For $x,y,z\in\{1,2\}$, we have
$(x,y,z)\in\theta^{\Bs_{\mathbb L}}$ if and only if the set
$R_{1,3}(x)\cap R^{-1}_{3,2}(y)\cap R_{3,3}(z)$ is non-empty. Intersecting the
sets displayed above gives
\[
\begin{array}{c|cccccccc}
(x,y,z) & (1,1,1) & (1,1,2) & (1,2,1) & (1,2,2) & (2,1,1) & (2,1,2) & (2,2,1) & (2,2,2)\\\hline
R_{1,3}(x)\cap R^{-1}_{3,2}(y)\cap R_{3,3}(z) & \varnothing & \{5,6\} & \{1\} & \{5,6\} & \{3,4\} & \{2\} & \{3,4\} & \varnothing
\end{array}
\]
This set is empty exactly when $x=y=z$, so $\theta^{\Bs_{\mathbb L}}=N$.

Hence $N$ is pp-definable in $\Bs_{\mathbb L}$. By
Theorem~\ref{thm:pp-reduction}, $\mathsf{CSP}(\mathcal N)$ reduces to
$\mathsf{CSP}(\Bs_{\mathbb L})$ in polynomial time. Therefore
$\mathsf{CSP}(\Bs_{\mathbb L})$ is NP-hard, and hence NP-complete.
\end{proof}

% Our goal in this section is to compare the occurrence of synchronous and $k$-partite quantum advantage, and to derive conditions under which $k$-partite quantum advantage can occur. 

\section{Synchronous vs $k$-partite advantage}
\label{sec:sync-vs-k-partite}

A natural question is to compare the $k$-partite definition of quantum advantage with the pre-existing synchronous notion given in definition \ref{def:synch-advantage}. The main result of this section separates the occurrence of these two forms of advantage by considering two well-known instances of $\mathsf{CSP}(\mathcal{XOR}_3)$, namely, the magic square and GHZ instances. Consider the signature $\sigma$ with two ternary symbols $R_{0}$ and $R_{1}$. We describe two structures over this signature:
\begin{itemize}
\item The structure $\mathcal{MS} \in \Rsig$ has universe $\{a_1, a_2, a_3, a_4, a_5, a_6, a_7, a_8, a_9\}$ and relations:
$$R_{0}^{\mathcal{MS}} = \{(a_1, a_2, a_3), (a_4, a_5, a_6), (a_7, a_8, a_9), (a_1, a_4, a_7), (a_2, a_5, a_8)\},$$
$$R_{1}^{\mathcal{MS}} = \{(a_3, a_6, a_9)\}.$$

\item The structure $\mathcal{GHZ} \in \Rsig$ has universe
$\{a_0, a_1, b_0, b_1, c_0, c_1\}$ and relations:
$$R_{0}^{\mathcal{GHZ}} = \{(a_0, b_0, c_0)\},$$
$$R_{1}^{\mathcal{GHZ}} = \{(a_0, b_1, c_1),\ (a_1, b_0, c_1),\ (a_1, b_1, c_0)\}.$$
\end{itemize}

$G^s(\mathcal{MS}, \mathcal{XOR}_3)$ is the Mermin-Peres magic square game \cite{mermin1993hidden,peres1990incompatible}. Under the colouring $p:G(\mathcal{GHZ}) \rightarrow K_3$ given by $
a_i \mapsto 1, b_i \mapsto 2, c_i \mapsto 3$, $G_p(\mathcal{GHZ}, \mathcal{XOR}_3)$ is the GHZ game \cite{greenberger1989going}.

\begin{theorem}
\label{thm:ms-ghz-separation}
The following holds:
\begin{itemize}
        \item $(\mathcal{GHZ}, \mathcal{XOR}_3)$ admits $3$-partite  $t$-advantage but not synchronous $co$-advantage.
        \item $(\mathcal{MS}, \mathcal{XOR}_3)$ admits synchronous  $t$-advantage but not $k$-partite $co$-advantage.
    \end{itemize}
\end{theorem}

Part of the proof of this theorem relies on a result of Arkhipov. 

\begin{definition}
\label{def:incidence}
A structure $\As \in \Rsig$ over the signature $\sigma = \{R_0, R_1\}$ is an \emph{incidence structure} if every $a \in A$ occurs
in exactly two of the tuple slots of $R^{\As}_0 \cup R^{\As}_1$.  Its
\emph{incidence multigraph} $Z_{\As}$ has one vertex for each tuple
$\vec a \in R^{\As}_0 \cup R^{\As}_1$ and one edge for each $a \in A$, joining the two tuples in which $a$ occurs. The \emph{parity} of $\As$ is
$\mathrm{par}(\As) := \bigl|R^{\As}_1\bigr| \bmod 2 \in \{0,1\}$.
\end{definition}
    
\begin{theorem}[\cite{arkhipov2012extending, paddock2023Arkhipov}]
\label{thm:arkhipov}
Let $\As$ be an incidence structure with a connected incidence multigraph. $(\As, \mathcal{XOR}_3)$ admits synchronous advantage if and only if
$\mathrm{par}(\As) = 1$ and $Z_{\As}$ is nonplanar.
\end{theorem}
\begin{proposition}
\label{prop:ghz-no-advantage}
$(\mathcal{GHZ}, \mathcal{XOR}_3)$ does not admit synchronous $co$-advantage.
\end{proposition}
\begin{proof}
$\mathcal{GHZ}$ is readily seen to be an incidence structure. Write $t_1 := (a_0, b_0, c_0)$ for the unique tuple of
$R^{\mathcal{GHZ}}_0$ and $t_2 := (a_0, b_1, c_1)$, $t_3 := (a_1, b_0, c_1)$,
$t_4 := (a_1, b_1, c_0)$ for the three tuples of $R^{\mathcal{GHZ}}_1$.  Each
element of the universe occurs in exactly two of these as required.

Moreover, $\mathrm{par}(\mathcal{GHZ}) = 1$ and $Z_{\mathcal{GHZ}} \cong K_4$ since each variable joins two distinct pairs from $\{t_1, t_2, t_3, t_4\}$ together. Hence $Z_{\mathcal{GHZ}}$ is planar and $(\mathcal{GHZ}, \mathcal{XOR}_3)$ admits no synchronous advantage.
\end{proof}

Our next aim is to show that $(\mathcal{MS}, \mathcal{XOR}_3)$ does not admit
$k$-partite $co$-advantage. For the case where $k=3$ this result follows from the work in \cite{ito2008generalized}. Here we present a novel proof which works for all values of $k$. Our technique can be viewed as a state-dependent variant of the substitution method attributed to Speelman in \cite{cleve2014characterization}, and may be of independent interest.

Throughout this section an \emph{observable} in a
unital $C^*$-algebra is an element $X$ with $X = X^{*}$ and $X^2 = I$, where we
abbreviate $I := I_{\mathbf{H}}$. We require several lemmas.

\begin{lemma}
\label{lem:kpartite-observable}
Let $\As \in \Rsig$ be a structure over $\sigma = \{R_0, R_1\}$ all of whose
tuples have pairwise distinct entries, let $p: G(\As) \rightarrow K_k$ be a
proper colouring, and let $\mathfrak{C} = (\mathbf{H}; \mathfrak{C}_1, \dots,
\mathfrak{C}_k; \psi)$ be a $k$-partite $C^*$-probability space.  Then
$(\As, p) \rightarrow Q^{co}_{\mathfrak{C}}(\mathcal{XOR}_3^{\cfk}, \pi_2)$ if and
only if there are observables $X_a \in \mathfrak{C}_{p(a)}$, one for each
$a \in A$, such that
\[
  X_{ a_1} X_{ a_2} X_{ a_3}\, \psi = (-1)^{\beta}\psi
  \qquad\text{for every } \beta \in \{0,1\} \text{ and every }
  \vec a \in R^{\As}_{\beta}.
\]
\end{lemma}

\begin{nestedproof}
Since $\pi_2^{-1}(\alpha) = \{0,1\} \times \{\alpha\}$, a Kleisli morphism
assigns to each $a \in A$ a PVM $P(a) \in \PVM(\mathfrak{C}_{p(a)},
\pi_2^{-1}(p(a)))$, which we identify with an element of
$\PVM(\mathfrak{C}_{p(a)}, \{0,1\})$.  The assignments $P \mapsto P(0) - P(1)$
and $X \mapsto \bigl(y \mapsto \tfrac12(I + (-1)^y X)\bigr)$ are mutually inverse
bijections between $\PVM(\mathfrak{C}_\alpha, \{0,1\})$ and the observables in
$\mathfrak{C}_\alpha$; put $X_a := P(a)(0) - P(a)(1)$.

Fix $\beta$ and $\vec a \in R^{\As}_\beta$.  The entries of $\vec a$ are
pairwise distinct, hence pairwise adjacent in $G(\As)$, hence assigned pairwise
distinct colours by $p$; so the operators $X_{a_i}$ lie in pairwise
commuting subalgebras and in particular commute.  By
Lemma~\ref{lemma:PVM-commeasurability} the family
$\Pi(\vec y) := P(\vec a_1)(y_1)P(\vec a_2)(y_2)P(\vec a_3)(y_3)$ is a PVM
indexed by $\{0,1\}^3$.  Setting $\Sigma_\gamma := \sum_{y_1 \oplus y_2 \oplus
y_3 = \gamma} \Pi(\vec y)$ we have $\Sigma_0 + \Sigma_1 = I$ and
\[
  X_{a_1} X_{a_2} X_{a_3}
  = \prod_{i \in [3]}\bigl(P(\vec a_i)(0) - P(\vec a_i)(1)\bigr)
  = \sum_{\vec y} (-1)^{y_1 \oplus y_2 \oplus y_3}\, \Pi(\vec y)
  = \Sigma_0 - \Sigma_1 .
\]
The relation preservation condition of Definition~\ref{def:sd-qhom} says that
$\Pi(\vec y)\psi = \pmb{0}$ for every $\vec y \notin R^{\mathcal{XOR}_3}_\beta$,
that is, for every $\vec y$ with $y_1 \oplus y_2 \oplus y_3 \neq \beta$.  If this
holds then $\Sigma_{1-\beta}\psi = \sum \Pi(\vec y)\psi = \pmb{0}$, the sum being
over exactly those $\vec y$; conversely if $\Sigma_{1-\beta}\psi = \pmb{0}$ then
$\|\Pi(\vec y)\psi\|^2 = \psi^\dagger \Pi(\vec y)\psi \leq \psi^\dagger
\Sigma_{1-\beta}\psi = 0$ for each such $\vec y$, since $\Pi(\vec y) \leq
\Sigma_{1-\beta}$.  Finally $\Sigma_{1-\beta}\psi = \pmb{0}$ holds if and only if
$\Sigma_\beta \psi = \psi$, if and only if $(\Sigma_0 - \Sigma_1)\psi =
(-1)^\beta \psi$.
\end{nestedproof}

We now specialise to $\mathcal{MS}$.  Write $a_{ij} := a_{3(i-1)+j}$ for
$i, j \in [3]$, so that the six tuples of $R^{\mathcal{MS}}_0 \cup
R^{\mathcal{MS}}_1$ are the three rows $(a_{i1}, a_{i2}, a_{i3})$ and the three
columns $(a_{1j}, a_{2j}, a_{3j})$, the third column being the unique tuple of
$R^{\mathcal{MS}}_1$.  Given observables $X_{ij} := X_{a_{ij}}$ as in
Lemma~\ref{lem:kpartite-observable}, put
\[
  \rho_i := X_{i1}X_{i2}X_{i3},
  \qquad
  \kappa_j := X_{1j}X_{2j}X_{3j},
\]
which are unambiguous since the three factors of each have distinct colours and
therefore commute. For $\theta\in S_3$, where $S_3$ denotes the symmetric group on $[3]$, set
\[
M_\theta
:=
\rho_1\kappa_{\theta(1)}
\rho_2\kappa_{\theta(2)}
\rho_3\kappa_{\theta(3)}.
\]

Two cells of the same row, or of the same column, are adjacent in
$G(\mathcal{MS})$, so each colour class of $p$ contains at most one cell from
each row and at most one from each column.  We call a set of cells with this
property a \emph{partial permutation}, and for $\pi \in S_3$ we write
$S_\pi := \{a_{i\pi(i)} : i \in [3]\}$; every partial permutation of size at
least two is contained in a unique $S_\pi$, since two cells with distinct rows
and distinct columns determine the third.

\begin{lemma}
\label{lem:word}
There is a $\theta \in S_3$, depending only on $p$, with $M_\theta = I$.
\end{lemma}

\begin{nestedproof}
Since the colour classes partition the nine cells, at most four colour classes have size at least two. Each such class is contained in a unique $S_\pi$, while $|S_3|=6$. Hence we may choose $\pi^*\in S_3$ such that no colour class of size at least two is contained in $S_{\pi^*}$. Set $\theta:=\pi^*(132).$

$M_\theta$ is a product of eighteen operators,
one for each of the six slots $\rho_1, \kappa_{\theta(1)},
\rho_2, \kappa_{\theta(2)}, \rho_3, \kappa_{\theta(3)}$ and each of the three
cells occurring in that slot.  Operators of distinct colours commute, so sorting
the eighteen factors by colour while preserving the relative order within each
colour, which is possible by finitely many transpositions of commuting
factors, gives
\[
  M_\theta = \prod_{\alpha \in [k]} M^{(\alpha)}_\theta ,
\]
where $M^{(\alpha)}_\theta$ is the product, in slot order, of the colour-$\alpha$ operators occurring in $M_\theta$. Fix a colour $\alpha$ and write
\[
S:=p^{-1}(\alpha).
\]
Since no two cells of $S$ lie in the same row or column, there exists
$\pi\in S_3$ such that
\[
S\subseteq S_\pi.
\]
Moreover, we may choose $\pi\neq\pi^*$. Indeed, if $|S|\geq 2$, the
permutation $\pi$ containing $S$ is unique and differs from $\pi^*$ by the
choice of $\pi^*$. If $|S|=1$, then $S$ is contained in two distinct
sets $S_\pi$, so at least one of the corresponding permutations differs
from $\pi^*$.

Write
\[
Y_i:=X_{i,\pi(i)}
\qquad (i\in[3]),
\]
and set
\[
\phi:=\theta^{-1}\pi.
\]
Consider first the word obtained by retaining all three cells of $S_\pi$.
The row factor $\rho_i$ contributes $Y_i$. The column factor
$\kappa_{\theta(t)}$ contributes the unique cell of $S_\pi$ lying in
column $\theta(t)$, namely
\[
a_{\pi^{-1}(\theta(t)),\,\theta(t)},
\]
and hence contributes
\[
Y_{\pi^{-1}(\theta(t))}
=
Y_{\phi^{-1}(t)}.
\]
Thus the corresponding six-letter word is
\begin{equation}
\label{eq:ms-word}
W_\phi
:=
Y_1Y_{\phi^{-1}(1)}
Y_2Y_{\phi^{-1}(2)}
Y_3Y_{\phi^{-1}(3)}.
\end{equation}

Since $\theta=\pi^*(132)$,
\[
\phi
=
\theta^{-1}\pi
=
(123)(\pi^*)^{-1}\pi.
\]
As $\pi\neq\pi^*$, we have $\phi\neq(123)$. Hence $\phi$ is one of the
remaining five elements of $S_3$, for which
\[
\begin{array}{rcl}
\phi=\id
&:&
W_\phi=Y_1Y_1\,Y_2Y_2\,Y_3Y_3=I,\\[2pt]
\phi=(12)
&:&
W_\phi=Y_1Y_2\,Y_2Y_1\,Y_3Y_3=I,\\[2pt]
\phi=(13)
&:&
W_\phi=Y_1Y_3\,Y_2Y_2\,Y_3Y_1=I,\\[2pt]
\phi=(23)
&:&
W_\phi=Y_1Y_1\,Y_2Y_3\,Y_3Y_2=I,\\[2pt]
\phi=(132)
&:&
W_\phi=Y_1Y_2\,Y_2Y_3\,Y_3Y_1=I.
\end{array}
\]
In each case the word reduces to $I$ by cancelling adjacent equal
factors, using only $Y_i^2=I$.

The actual colour class $S$ need not contain all three cells of $S_\pi$.
The colour-$\alpha$ subword $M_\theta^{(\alpha)}$ is obtained from
$W_\phi$ by deleting both occurrences of $Y_i$ whenever
$a_{i,\pi(i)}\notin S$. Equivalently, define a homomorphism
\[
r_S:
\ast_{i=1}^3\mathbb Z_2
\longrightarrow
\ast_{a_{i,\pi(i)}\in S}\mathbb Z_2
\]
on the generators by
\[
r_S(Y_i)
=
\begin{cases}
Y_i, & a_{i,\pi(i)}\in S,\\
I,   & a_{i,\pi(i)}\notin S.
\end{cases}
\]
Then
\[
M_\theta^{(\alpha)}=r_S(W_\phi).
\]
Since $W_\phi=I$, it follows that
\[
M_\theta^{(\alpha)}
=
r_S(W_\phi)
=
r_S(I)
=
I.
\]

Thus $M_\theta^{(\alpha)}=I$ for every colour $\alpha$, and therefore
\[
M_\theta
=
M_\theta^{(1)}\cdots M_\theta^{(k)}
=
I.
\]

\end{nestedproof}

\begin{proposition}
\label{prop:ms-no-advantage}
$(\mathcal{MS}, \mathcal{XOR}_3)$ does not admit
$k$-partite $co$-advantage.
\end{proposition}

\begin{proof}
Suppose that $(\mathcal{MS}, p) \rightarrow Q^{co}_{\mathfrak{C}}
(\mathcal{XOR}_3^{\cfk}, \pi_2)$ for some $k$, some proper colouring $p$ and some
$k$-partite $C^*$-probability space $\mathfrak{C}$.  The tuples of
$\mathcal{MS}$ have pairwise distinct entries, so
Lemma~\ref{lem:kpartite-observable} supplies observables $X_{ij} \in
\mathfrak{C}_{p(a_{ij})}$ with
\[
  \rho_i\, \psi = \psi \quad (i \in [3]),
  \qquad
  \kappa_j\, \psi = (-1)^{[\,j = 3\,]}\psi \quad (j \in [3]).
\]
Let $\theta$ be as in Lemma~\ref{lem:word}.  Each of the six factors of
$M_\theta$ maps $\psi$ to a scalar multiple of $\psi$, so applying them from
right to left multiplies the scalars; since each of the six tuples of
$\mathcal{MS}$ occurs exactly once in $M_\theta$ and exactly one of them lies in
$R^{\mathcal{MS}}_1$,
\[
  M_\theta\, \psi = (-1)^{1}\psi = -\psi .
\]
But $M_\theta = I$ by Lemma~\ref{lem:word}, so $\psi = -\psi$ and hence
$\psi = \pmb{0}$, contradicting $\|\psi\| = 1$.  Therefore no perfect
$co$-strategy exists, for any $k$ and any proper colouring $p$.
\end{proof}

We can now prove the main theorem of this section. 

\begin{proof}[Proof of theorem \ref{thm:ms-ghz-separation}]
    The fact that $(\mathcal{GHZ}, \mathcal{XOR}_3)$ admits advantage is equivalent to the well-established result that the GHZ game has a perfect quantum strategy \cite{greenberger1989going} but no classical strategy. Similarly, $(\mathcal{MS}, \mathcal{XOR}_3)$ admits synchronous advantage because the magic square game exhibits this form of advantage. The remainder of the claim was shown in propositions \ref{prop:ghz-no-advantage} and \ref{prop:ms-no-advantage}.
\end{proof}

\section{Bounded width}
\label{sec:bounded-width}
Theorem~\ref{thm:ms-ghz-separation} separates two forms of quantum advantage at the level of instances. We now show that at the level of templates the two notions share a necessary condition. As mentioned in the introduction Ciardo has shown that a CSP template admitting synchronous quantum advantage must have unbounded width \cite{ciardo2024quantum}. We show that any template satisfying $k$-partite advantage must also satisfy this condition. It remains an open question whether a template exists which admits one form of advantage but not the other.

The width of $\mathsf{CSP(\Bs)}$ is a central notion in constraint satisfaction. It can be defined in many equivalent ways. One such definition says that $\mathsf{CSP(\Bs)}$ has width $k$ iff the $k$-consistency algorithm \cite{feder1998conjecture} successfully solves it. \cite{barto2016robust} shows that a template has bounded width iff the SDP relaxation correctly solves it. We now define this concept.

\begin{definition}
\label{def:sdp}
Let $\As,\Bs$ be finite $\sigma$-structures. The system $\mathrm{SDP}(\As,\Bs)$ has as its
variables a vector $u_{a,b}$ for each $a\in A$ and $b\in B$, a scalar $v_{R,\vec a,\vec b}$
for each $R\in\sigma$, $\vec a\in R^\As$ and $\vec b\in R^\Bs$, and a distinguished vector
$u_0$ with $\|u_0\|=1$; the vectors range over a finite-dimensional complex inner product space
and the scalars over $\mathbb R$. The constraints are
\begin{align}
\sum_{\vec b\in R^\Bs} v_{R,\vec a,\vec b} &= 1, \label{eq:sdp-norm}\\[-2pt]
v_{R,\vec a,\vec b} &\geq 0, \label{eq:sdp-pos}\\[-2pt]
\langle u_{a_i,b},\,u_0\rangle
&= \sum_{\vec b\in R^\Bs,\ b_i=b} v_{R,\vec a,\vec b}, \label{eq:sdp-first}\\[-2pt]
\langle u_{a_i,b},\,u_{a_j,b'}\rangle
&= \sum_{\vec b\in R^\Bs,\ b_i=b,\ b_j=b'} v_{R,\vec a,\vec b}, \label{eq:sdp-second}
\end{align}
where $R$ ranges over $\sigma$, $\vec a$ over $R^\As$, the indices $i,j$ over $[\ar(R)]$, and
$b,b'$ over $B$. We say that $\mathrm{SDP}$ \emph{solves} $\mathsf{CSP(\Bs)}$ if feasibility of
$\mathrm{SDP}(\As,\Bs)$ implies $\As\rightarrow\Bs$, for every $\As$.
\end{definition}
\begin{theorem}
\label{thm:kpartite-sdp}
If $(\As,p)\rightarrow Q^{co}_{\mathfrak{C}}(\Bs^{\cfk},\pi_2)$, then $\mathrm{SDP}(\As,\Bs)$ is feasible by setting

$$
u_0:=\psi,
\qquad
u_{a,b}:=P(a)(b)\,\psi,
\qquad
v_{R,\vec a,\vec b}:=\psi^\dagger P(\vec a) (\vec b) \psi.
$$
\end{theorem}

\begin{proof}
\emph{Constraint \eqref{eq:sdp-norm}.}
$$
\sum_{\vec b\in R^\Bs}v_{R,\vec a,\vec b}
=\sum_{\vec b\in B^n}\psi^\dagger P(\vec a)(\vec b)\psi
=\psi^\dagger\Bigl(\sum_{\vec b\in B^n}P(\vec a)(\vec b)\Bigr)\psi
=\psi^\dagger\psi=1.
$$

\emph{Constraint \eqref{eq:sdp-pos}.} The operator
$P(\vec a)(\vec b)$ is an orthogonal projection, so
$v_{R,\vec a,\vec b}=\psi^\dagger P(\vec a)(\vec b)\psi=\|P(\vec a)(\vec b)\psi\|^2\geq 0$; in particular each
$v_{R,\vec a,\vec b}$ is a non-negative real number.

\emph{Constraint \eqref{eq:sdp-first}.} Using the marginalisation property of the joint PVM $P(\vec a)$ and self-adjointness we obtain

$$\sum_{\substack{\vec b\in R^\Bs\\ b_i=b}}
v_{R,\vec a,\vec b}
=
\sum_{\substack{\vec b\in B^n\\ b_i=b}}
\psi^\dagger P(\vec a)(\vec b)\psi
=
\psi^\dagger P(a_i)(b)\psi
=
\bigl(P(a_i)(b)\psi\bigr)^\dagger\psi
=
\langle u_{a_i,b},u_0\rangle.
$$

\emph{Constraint \eqref{eq:sdp-second}.}
By relation preservation, terms with $\vec b\notin R^\Bs$ vanish on $\psi$. Hence, using the two-coordinate marginalisation property of the joint PVM $P(\vec a)$,
\[
\begin{aligned}
\sum_{\substack{\vec b\in R^\Bs\\ b_i=b,\ b_j=b'}}
v_{R,\vec a,\vec b}
&=
\sum_{\substack{\vec b\in B^n\\ b_i=b,\ b_j=b'}}
\psi^\dagger P(\vec a)(\vec b)\psi\\
&=
\psi^\dagger P(a_i)(b)P(a_j)(b')\psi\\
&=
\bigl(P(a_i)(b)\psi\bigr)^\dagger
\bigl(P(a_j)(b')\psi\bigr)\\
&=
\langle u_{a_i,b},u_{a_j,b'}\rangle.
\end{aligned}
\]
% By Lemma~\ref{lem:intra-constraint} the operator $P(a_i)(b)P(a_j)(b')$ is a product of two
% commuting projections, hence itself a projection, so this quantity equals
% $\|P(a_i)(b)P(a_j)(b')\psi\|^2\geq 0$ and is in particular real. 

% Three degenerate cases are worth
% recording, since they constrain the assignment most tightly. If $i=j$ and $b\neq b'$ the sum is
% empty, and indeed $\langle u_{a_i,b},u_{a_i,b'}\rangle=\psi^\dagger P(a_i)(b)P(a_i)(b')\psi=0$: the
% vectors attached to a single element of $A$ are pairwise orthogonal, as the system demands, and
% this is precisely projectivity of $P(a_i)$. If $i=j$ and $b=b'$ the two sides reduce to the two
% sides of \eqref{eq:sdp-first}. If $i\neq j$ but $a_i=a_j$, both sides equal
% $\delta_{bb'}\psi^\dagger P(a_i)(b)\psi$; here the tuples $\vec b\in R^\Bs$ with $b_i\neq b_j$
% contribute nothing because $\Pi(\vec b)=\pmb 0$ for inconsistent $\vec b$, and not because they are
% excluded from $R^\Bs$, which they need not be.

\emph{Finite dimensionality.} Let $V\subseteq\HH$ be the complex-linear span of
$\{\psi\}\cup\{P(a)(b)\psi: a\in A,\ b\in B\}$, so that $\dim_{\mathbb C}V\leq 1+|A||B|<\infty$. Then all the vectors above lie in $V$ as required.
\end{proof}

Feasibility of $\mathrm{SDP}(\As,\Bs)$ is thus a necessary condition for a perfect $k$-partite $co$-strategy. As every $t$-strategy is a $co$-strategy this makes it necessary for those strategies as well. Since a CSP template has bounded width iff SDP solves it we obtain the following corollary.

\begin{corollary}
\label{cor:shared-condition}
If $\mathsf{CSP(\Bs)}$ admits $k$-partite $u$-advantage for some $k$ and some $u\in\{t,co\}$, then $\mathsf{CSP(\Bs)}$ has unbounded width.\end{corollary}

\begin{proof}
    Suppose $\mathsf{CSP}(\Bs)$ admits $k$-partite $u$-advantage. Then there exists $\As\nrightarrow\Bs$ admitting a perfect $k$-partite $u$-strategy. By Theorem~\ref{thm:kpartite-sdp}, and since every $t$-strategy is a $co$-strategy, $\mathrm{SDP}(\As,\Bs)$ is feasible. Hence $\mathrm{SDP}$ does not solve $\mathsf{CSP}(\Bs)$, so $\mathsf{CSP}(\Bs)$ has unbounded width. 
\end{proof}

\section{Outlook}

Our work suggests several directions for further study. We list three possibilities below:

\begin{itemize}
    \item We have focused on pseudotelepathic quantum advantage. Can our framework be extended to explain probabilistic quantum advantage, as exemplified by the CHSH game \cite{clauser1969proposed}? What about logical quantum advantage, as exemplified by Hardy's paradox \cite{hardy1992quantum,hardy1993nonlocal}?
    \item In \cite{ciardo2024quantum, ciardo2026quantumpolymorphisms, bulatov2025satisfiability} tools from the universal algebraic approach to CSPs, such as minions, gadget reductions, and polymorphisms are used to explore the occurrence of synchronous quantum advantage, as well as the complexity of synchronous quantum CSP viewed as a decision problem. Can we use these ideas to study the $k$-partite variations of these concepts? 
    
%     As a first tool in this direction, we show, following results from \cite{connolly2022game, barto2021algebraic} that the task of identifying quantum advantage can be reduced to proving the non-existence of certain homomorphisms in $\ksig$.

% \begin{theorem}
%     The following are equivalent:
%     \begin{enumerate}
%         \item $\mathsf{CSP(\Bs)}$ admits $u$-quantum advantage.
%         \item There does not exists a morphism $Q^u_g (\Bs^{\cfk}, \pi_2) \rightarrow (\Bs^{\cfk}, \pi_2)$.
%     \end{enumerate}
% \end{theorem}

% \begin{proof}
%     We prove the contrapositive. 
    
%     Assume that $\mathsf{k-CSP^c(\Bs)}$ does not admit $u$-quantum advantage. This is equivalent to saying
%     $$(\As, p) \rightarrow Q^u_g (\Bs^{\cfk}, \pi_2) \iff (\As, p) \rightarrow (\Bs^{\cfk}, \pi_2)$$ for any $(\As, p) \in \ksig.$ Take $(\As, p) = Q^u_g (\Bs, \pi_2)$. The identity morphism gives a map $Q^u_g (\Bs, \pi_2) \rightarrow Q^u_g (\Bs^{\cfk}, \pi_2)$. Via the above this implies the existence of a map $Q^u_g (\Bs^{\cfk}, \pi_2) \rightarrow (\Bs^{\cfk}, \pi_2).$

%     Conversely, assume $f:Q^u_g (\Bs^{\cfk}, \pi_2) \rightarrow (\Bs^{\cfk}, \pi_2)$ exists. If $(\As, p) \rightarrow Q^u_g (\Bs^{\cfk}, \pi_2)$ then composing this map with $f$ gives $(\As, p) \rightarrow (\Bs^{\cfk}, \pi_2)$. Conversely, if $(\As, p) \rightarrow (\Bs^{\cfk}, \pi_2)$ composing with the monad unit gives $(\As, p) \rightarrow Q^u_1 (\Bs^{\cfk}, \pi_2)$.
% \end{proof}

\item In the synchronous setting a quantum isomorphism game on graphs was introduced in \cite{atserias2019quantum} and has been extensively studied since \cite{mancinska2020quantum, lupini2020nonlocal, nigam2024npa, musto2018compositional}. This game was generalised to relational structures in \cite{karamlou2025quantum}. Is there a corresponding notion of $k$-partite quantum isomorphism?

\end{itemize}

\bibliographystyle{acm}
\bibliography{ref}

\section*{Use of AI tools}
The author discloses the use of AI tools, specifically Claude (Anthropic) and ChatGPT (OpenAI) in preparing this submission. These tools were used to generate drafts of several proofs, which were then verified, and edited heavily by the author to improve clarity and correctness. The models assisted with drafting parts of the exposition in a similar manner.
Every statement in this document has been checked by the author, who take full responsibility for the correctness and for the content of the submission as a whole.

\end{document}